\documentclass[12pt,letterpaper]{article}
\usepackage[utf8]{inputenc}
\usepackage[english]{babel}
\usepackage[vmargin=0.6in,hmargin={1in,1in},headsep=0.3in,headheight=1in]{geometry}
\usepackage{subfiles}
\usepackage{tikz}
\usepackage{fix-cm}
\usepackage{enumitem}
\usepackage{multicol}
\usepackage{amsmath, amsthm, amssymb, amsfonts,mathrsfs}
\usepackage{paracol}
\usepackage{graphicx}

\usepackage[normalem]{ulem}

\usepackage[bottom,multiple]{footmisc}  

\stepcounter{footnote}

\usepackage{manyfoot}
\DeclareNewFootnote{A}
\renewcommand{\footnote}[1]{\footnoteA{#1}}

\allowdisplaybreaks
\newcounter{taggedEquations}
\let\OldTag\tag
\renewcommand*{\tag}[1]{\stepcounter{taggedEquations}\OldTag{#1}}
\usepackage{array}
\usepackage{tabularx}

\usepackage{relsize}
\usepackage{xcolor}
\usepackage{lastpage}
\usepackage{blindtext}
\usepackage[numbers,sort&compress]{natbib}
\usepackage[linktocpage=true]{hyperref}
\hypersetup{
	colorlinks,
	citecolor=blue,
	filecolor=black,
	linkcolor=blue,
	urlcolor=blue
}
\usepackage{url}

\usepackage{caption}
\usepackage{subcaption}
\usepackage{wrapfig}
\usepackage{indentfirst}
\usepackage{hyperref}
\usepackage{soul}
\usepackage{bm}

\usepackage{empheq}
\usepackage{authblk}
\setlist{parsep=0pt,listparindent=\parindent}
\setlist[itemize]{noitemsep, topsep=0pt}
\setlist[enumerate]{noitemsep, topsep=0pt}
\setlist{parsep=0pt,listparindent=\parindent}

\stepcounter{footnote}
\graphicspath{{Images/}{../Images/}}
\usepackage{graphicx}
\graphicspath{ {./images/} }
\usepackage{pgfplots}
\usepackage{extarrows}
\usepackage{mathtools}
\usepackage{tikz}
\usetikzlibrary{decorations.pathmorphing,patterns}
\usetikzlibrary{decorations.pathreplacing,angles,quotes}
\usetikzlibrary{patterns}
\usetikzlibrary{shapes.arrows, fadings}
\usepackage{xfrac} 
\usepackage{relsize}
\usepackage[most]{tcolorbox}
\tikzfading[name=fade down,
  top color=transparent!0, bottom color=transparent!100]
\tikzfading[name=fade up,
  bottom color=transparent!0, top color=transparent!100]

\usepackage{pgfplots}
\pgfplotsset{compat=1.18}
\usetikzlibrary{external}

\usepackage{physics}

\usepackage{enumitem}

\def\beq{\begin{eqnarray}}  
\def\eeq{\end{eqnarray}}

\begin{document}

\title{\Large\textbf{An Intrinsic Coordinate Reference Frame Procedure I: Tensorial Canonical Weyl Scalars}}
\author[1,2,3]{{\large{C.~K.~Watson\footnote{mail to: \em\texttt{\href{cooper_watson@Baylor.edu}{Cooper\_Watson@Baylor.edu}}}}}}
\author[2,3,4]{\large{W.~Julius}}
\author[2,3]{\large{P.~Brown}}
\author[5]{\large{D.~Salisbury}}
\author[2,3]{\large{G.~B.~Cleaver}}

\affil[1]{\emph{Institute for Quantum Science \& Engineering (IQSE), Texas A\&M University, College Station, Texas 77843, USA}}
\vspace{1 cm}
\affil[2]{\emph{Early Universe, Cosmology and Strings (EUCOS) Group, Center for Astrophysics, Space Physics and Engineering Research (CASPER)}}
\vspace{1 cm}
\affil[3]{\emph{Department of Physics, Baylor University,  Waco, TX 76798, USA} }
\vspace{1 cm}
\affil[4]{\emph{College of Science and Engineering (COSE), St. Cloud State University, St. Cloud, MN 56301, USA}}
\vspace{0.25cm}
\affil[5]{\emph{Austin College, 900 North Grand Avenue, Sherman, TX 75090, USA}}
\vspace{0.25cm}

\date{\today}  
\maketitle  
\begin{abstract}   

The canonical quantization of gravity in general relativity is greatly simplified by the artificial decomposition of space-time into a 3+1 formalism. Such a simplification appears to come at the cost of general covariance. This quantization procedure requires tangential and perpendicular infinitesimal diffeomorphisms generated by the symmetry group under the Legendre transformation of the given action. This gauge generator, along with the fact that Weyl curvature scalars may act as ``intrinsic coordinates" (or a dynamical reference frame) which depend only on the spatial metric $g_{ab}$ and the conjugate momenta $p^{cd}$, allow for an alternative approach to canonical quantization of gravity. In this paper we present the tensorial solution of the set of Weyl scalars in terms of canonical phase-space variables.

\end{abstract} 

\par

\pagebreak

\section{Introduction}

Since its discovery in the early 1900's, the General Theory of Relativity (GR) has suffered from an embarrassment of riches \cite{einstein_1905,einstein_1915}. Within a decade, GR was already in conflict with the emerging field of Quantum Mechanics (QM) \cite{Solvay_1927}. Substantial progress was made over the subsequent 50 years, culminating in Dirac's theory of gravitation in a Hamiltonian formalism \cite{dirac_1958b}. Dirac inadvertently implied that infinitesimal diffeomorphisms must incorporate a compulsory metric field dependence and decompose spacetime into a (3+1) formalism \cite{salisbury_2020}. This reformulation, popularized by Arnowitt, Deser, and Misner (ADM), defined the lapse and shift functions ($N,N^a$) to describe the arbitrary (3+1) decomposition of the spatial-hypersurface and the time parameter \cite{arnowitt_gravitation_1962,dirac_1958b,salisbury_2020}. This was most often used in the context of J. A. Wheeler's geometrodynamical program to simply describe the evolution of solutions to Einstein's field equations \cite{gravitation}. Although the use of ADM formalism gained traction, the idea of defining the lapse and shift functions as phase-space variables was abandoned \cite{salisbury_2020}. Unknown to many, the canonical quantization approach was hindered by Wheeler's use of three-geometries. Wheeler's method focused solely on the spatial covariance and failed to incorporate the full four-dimensional (4D) diffeomorphism symmetry. In the late 1950's, those in the school of Wheeler's geometrodynamics were able to argue failures of similar theories, unaware that their criticism also applied to their own approach\footnote{We believe this is most likely caused by Bergmann's work being deeply rooted in the passive diffeomorphism view.} {\cite{bergmann_1959}}. 


Any alternate approach to quantize gravity requires promoting some set of observables and operators to describe the degrees of freedom of the respective phase-space. The use of local observables or local quantum operators requires some ``intrinsic coordinate system" (ICS) or a specific dynamical reference frame (DRF) defining the gravitational interactions of the bodies that form the reference frame. We refer to this ICS (or specific DRF) as an ``intrinsic coordinate reference frame" (ICRF) for the remainder of the paper. This is a necessary clarification as a quantum reference frame differs from a classical reference frame \cite{kabel_2024}. One candidate for an ICRF is the set of Weyl scalars as proposed by Komar in \cite{komar_construction_1958} where the spacetime geometry is described by the Weyl scalars. We make this choice since GR is a fundamentally local theory which is background independent. Here background independence means that the theory does not rely on an\textit{a priori} choice of geometry, but rather the geometry is determined by the dynamics of the theory itself \cite{chataignier_diss_2022}. This subtlety is often overlooked but has serious consequences. In our case it becomes increasingly problematic when discussing ``Geometrodynamics" in which the geometry, and therefore the coordinate labelings, \textit{are} dynamical. This is exemplified by quantum mechanics in \cite{rovelli_2014gauge}, where the dynamics of two equivalent, but isolated, systems differ when described by the dynamics of the two systems coupled. In other words: there are no external objects in a true relativistic theory, so the lack of a relationship between two distinct events fails to incorporate the full dynamical behavior. Consequently, it becomes compulsory that we recognize the underlying 4D diffeomorphism symmetry.


We adopt the Rosenfeld-Bergmann-Dirac (RBD) approach because of its ability to describe a total Hamiltonian by only primary and secondary first-class constraints $(P_{\mu}=0,~\mathcal{H}_{\mu}=0)$ where the constraints hold for all time given the initial conditions \cite{chataignier_diss_2022}. Working with the Hamilton-Jacobi formalism we use the Bergmann-Komar (BK) group, but \textit{enlarge} the group to include the lapse and shift in the configuration space. Doing so, we can relate the gauge symmetry and the concept of time as an evolutionary parameter \cite{pons_2005}.

In this paper we present the tensorial solution to the Weyl scalars in terms of the canonical phase-space variables. It is already well known that the Weyl scalars are a set of scalar polynomial coordinate invariants (SPCIs)\footnote{Traditionally we refer to scalar coordinate invariants simply as ``invariants" (SPIs) \cite{carminati1991algebraic}. We adopt the ``C" to distinguish coordinate and gauge invariants.} related to the original form of G\'eh\'enau and Debever coordinate invariants. We note that traditional coordinate invariant scalars are not gauge-invariant, whereas the Weyl scalars presented here \textit{are} gauge-invariant scalars and depend only on the spatial metric and conjugate momenta \cite{salisbury_cartan_2022,bk_1960, bergmann_1961,pons_revisiting_2009}. Due to the density of the topics discussed here, we assume the reader to be familiar with the current understanding of observables \cite{rovelli_1996,rovelli_2002b,pons_revisiting_2009,chataignier_construction_2020,chataignier_relational_2021,chataignier_diss_2022}, the intrinsic Hamilton-Jacobi approach \cite{de_leon_2013,salisbury_restoration_2016,rosenfeld_quantization,salisbury_2020,salisbury_history_2022,salisbury_cartan_2022}, canonical QG methods \cite{isham_1993}, and intrinsic coordinates \cite{pons_1997,pons_revisiting_2009,pons_gravitational_2009,salisbury_cartan_2022}.

For clarity we list the indexing convention used here. Lower case Greek indices ($\alpha,\beta$,...) refer to coordinates on the pseudo-Riemannian manifold $\mathcal{M}$ and take values from 0 to 3. Lower case Latin indices ($a,b,...$) take on the values of 1 to 3, indicating the spatial indices. Upper case Latin indices refer to internal indices ($I,J,...$) and run from 0 to 3 (or 1 to 4 for scalars). We reserve lower case Latin indices ($i,j,...$) for the internal spatial indices, which run from 1 to 3. The calligraphic capital Latin indices ($\mathcal{A},\mathcal{B},...$) refer to bivector indices and run from 1 to 6 for the symplectic group. 

Note that our tetrad vectors do not transform as 4-vectors since they are coupled to the time foliation \cite{pons_gauge_triad_2000,salisbury_2023}. We can return the metric tensor by using the internal metric
\begin{equation}
\begin{gathered}
    \eta_{IJ}=\eta^{IJ}=\text{diag}\left(-1,1,1,1\right) \notag
\end{gathered}
\end{equation}
and the relations
\begin{equation}
\begin{gathered}
    g_{\mu\nu}=\eta_{IJ}e^I_{\mu}e^J_{\nu}, \\
    g^{\mu\nu}=\eta^{IJ}E^{\mu}_{I}E^{\nu}_{J}. \notag \\
\end{gathered}
\end{equation}
We also must mention the definition of $\Bar{\delta}$. For a given field $F(x)$, let $F'(x')$ represent the field obtained under the infinitesimal coordinate transformation $x'^{\mu}=x^{\mu}-\epsilon^{\mu}$, then $\Bar{\delta}F:=F'(x')-F(x)$. This is the Lie derivative with respect to $\epsilon^{\mu}$.

The ADM Lagrangian transforms as a scalar density (weight +1) under the active diffeomorphism-induced field transformations $\left(\overline{\delta}\mathcal{L}_{ADM}{\equiv}\left(\mathcal{L}_{ADM}\epsilon^{\mu}\right)_{,\mu}\right)$. It was shown in \cite{salisbury_cartan_2022} that this property allows us to utilize Noether's second theorem \cite{noether_1918} so that the vanishing Rosenfeld-Noether charge, $\mathfrak{C}^0_{ADM}$, is simply the generator of infinitesimal gauge transformations, $G_{\xi}(x^0)$:
\begin{equation}
    {\int}d^3x~\mathfrak{C}^0_{ADM}={\int}d^3x~G_{\xi}(x^0). 
\end{equation}
We encounter an obstacle in which the traditional commutator Lie algebra, $\epsilon^{\mu}=\epsilon^{\mu}_{1,\nu}\epsilon^{\nu}_{2}-\epsilon^{\mu}_{2,\nu}\epsilon^{\nu}_{1}$, requires higher time-derivatives and so the Rosenfeld-Noether charge is not projectable to the canonical phase-space under the Legendre transformations \cite{bergmann_1962,bergmann_alternative_1976,bergmann_fading_1979,salisbury_cartan_2022,salisbury_2023}. This is because the diffeomorphism-induced gauge transformations \textit{are} projectable under the Legendre transformation if, and only if, the infinitesimal variations depend on the lapse and shift but not their time derivatives \cite{pons_1997,pons_gauge_yang_2000}. Instead, we are required to define the diffeomorphism transformations to be dependent on the metric:
\begin{equation}
    \epsilon^{\mu} := \xi^{\mu}=\delta^{\mu}_{a}\xi^a+n^{\mu}\xi^0, \label{epsilon}
\end{equation}
where $n^{\mu}$ is the normal to the constant time hypersurface: 
\begin{equation}
    n^{\mu}=-g^{0\mu}\left(-g^{00}\right)^{-1/2}=\left(N^{-1},-N^{-1}N^a\right). \label{unit_normal} \\
\end{equation}
The corresponding commutator algebra is
\begin{equation}
\begin{gathered}
	{\xi}^{\mu}={\delta}^{\mu}_{a}\left({{\xi}^{a}_{1,b}}{{\xi}^{b}_{2}}-{{\xi}^{a}_{2,b}}{{\xi}^{b}_{1}}\right)
    +e^{ab}\left({{\xi}^0_{1}}{{\xi}^0_{2,b}}-{{\xi}^0_{2}}{{\xi}^0_{1,b}}\right)+n^{\mu}\left({{\xi}^a_{2}}{{\xi}^0_{1,a}}-{{\xi}^a_{1}}{{\xi}^0_{2,a}}\right),  \label{commutator_algebra}
\end{gathered}
\end{equation}
where $e^{ab}$ is the inverse of the spatial metric $g_{ab}$ (we differ from ADM or traditional notation where $e^{ab}$ is often denoted as $\gamma^{ab}$ or ${}^{(3)}g^{ab}$ respectively). This was first derived by Bergmann when he noticed that there was no simple relationship between Dirac's phase-space (configuration space) and the fiber-bundle(s) over the spacetime manifold \cite{bergmann_fading_1979,salisbury_history_2022}. This result is derived explicitly in \cite{pons_1997}.

With this decomposition we have the complete vanishing phase-space generator of diffeomorphism-induced transformations
\begin{equation}
\begin{gathered}
    G_{\xi}\left(x^0\right)=P_{\mu}\dot{\xi}^{\mu}+\mathcal{H}_{\mu}\xi^{\mu}+P_0\left(\xi^0-N^a\xi^0_{,a}+N_{,a}\xi^a\right)\\
    +P_a\left(N_{,b}e^{ab}\xi^0_{,b}-N\xi^0_{,b}e^{ab}+N_{,a}\xi^a+N^a_{,b}\xi^b-N^b\xi^a_{,b}\right), \label{generator} \\
\end{gathered}
\end{equation}
as originally presented in \cite{salisbury_cartan_2022}. Here $P_{\mu}$ is the conjugate momenta of the shift. It is important to note that the complete generator depends on the term $\dot{\xi}^{\mu}$, and so the variations of the lapse and shift depend on time-derivatives of the diffeomorphisms. On the other hand, the variation of $g_{ab}$ and $p^{ab}$ under the infinitesimal transformations \eqref{epsilon} do not depend on the time derivative $\dot{\xi}$ so they are ``D-invariant"\footnote{Bergmann gave this name in honor of Dirac. It describes an object whose transformation under a diffeomorphism does not depend on the time derivative of the diffeomorphism \cite{bergmann_1962}.}. To be more precise, consider infinitesimal coordinate transformations
\begin{equation}
    x'^{\mu}=x^{\mu}-\epsilon^{\mu}, \label{diffeo}
\end{equation}
such that $\epsilon^{\mu}$ vanishes on the hypersurface $x^0$, but assume ${\partial\epsilon^{\mu}}/{\partial{x^0}}$ does not vanish on this hypersurface. Then a D-invariant is, by Bergmann's definition, a variable that remains invariant under this transformation. In other words - its variation does not depend on this time derivative. This is still fully covariant under active general coordinate transformations as long as we enlarge the phase-space to have the infinitesimal coordinate transformations dependent on the lapse and shift functions \cite{salisbury_cartan_2022}. 

To compute the Weyl scalars in terms of canonical phase-space variables we must cast the Weyl tensor into symplectic space with our enlarged BK group. Then we write the scalars in ADM and tetrad formalisms in terms of the lapse and shift functions. These two mathematical forms will each produce two scalars. 

\subsection{ADM Formalism}

Starting with the ADM Lagrangian we follow \cite{salisbury_2023}, where we use minus one-half the ADM Lagrangian to introduce the triad variables in the following section
\begin{equation}
\begin{gathered}
    \mathcal{L}=-\frac{1}{2}\mathcal{L}_{ADM}=-\frac{1}{2}Nt\left({}^{(3)}R+K_{ab}K^{ab}-\left({K^a}_a\right)^2\right) \\
    =-\frac{1}{2}Nt\left({}^{(3)}R+K_{ab}e^{ac}e^{bd}K_{cd}-\left(K_{ab}e^{ab}\right)^2\right).
\end{gathered}
\end{equation}
Here the extrinsic curvature $K_{ab}$ is given by
\begin{equation}
\begin{gathered}
    K_{ab}=\frac{1}{2N}\left(g_{ab,0}-N^cg_{ab,c}-g_{ca}{N^c}_{,b}-g_{cb}{N^c}_{,a}\right) \\
    =\frac{1}{2N}\left(g_{ab,0}-2g_{c(a}{N^c}_{|b)}\right).
\end{gathered}
\end{equation}
Additionally, we define
\begin{equation}
    t:=\sqrt{{}^{(3)}g},
\end{equation}
where ${}^{(3)}g$ is the determinant of the spatial metric $g_{ab}$. For completeness we also explicitly define the conjugate momenta
\begin{equation} \label{conjmomenta}
\begin{gathered}
    p^{ab}=\sqrt{{}^{(3)}g}\left(K^{ab}-{K^c}_ce^{ab}\right), \\
    p_{ab}=\sqrt{{}^{(3)}g}\left(K_{ab}-{K^c}_cg_{ab}\right).
\end{gathered}
\end{equation}
We make the distinction between the three-dimensional curvature scalar, ${}^{(3)}R$, and the four-dimensional curvature scalar ${}^{(4)}R$, where we also apply this notation for the traditional or commonplace (4D) Riemannian tensors. 

Since we want the canonical description (in phase-space), we will write the covariant (and contravariant) space-time metric(s) in terms of the lapse function $N$ and the shift vector $N^a$,
\begin{equation}
\begin{gathered}
    g_{\alpha\beta}=
        \begin{pmatrix} 
         -N^2+g_{ab}N^aN^b  & g_{ab}N^a \\
            g_{ab}N^b & g_{ab} \\
        \end{pmatrix}, ~~
    g^{\alpha\beta}=
        \begin{pmatrix} 
         -N^{-2}  & N^{-2}N^a \\
            N^{-2}N^a & e^{ab}-N^aN^bN^{-2} \\
        \end{pmatrix}.
\end{gathered}
\end{equation}
We note that the contravariant metric may be written as 
\begin{equation}
\begin{gathered}
    g^{\mu\nu}=e^{\mu\nu}-n^{\mu}n^{\nu}, \label{projection tensor} 
\end{gathered}
\end{equation}
where $e^{\mu\nu}$ is the projection tensor, and is one of the three original D-invariants from Dirac \cite{dirac_1958b}. We will also use the D-invariant:
\begin{equation}
    e^{0\mu}=0, \label{DI1}
\end{equation}
to work ``on-shell" of the hypersurface.
We write the Weyl tensor (conformal tensor)
\begin{equation}
\begin{gathered}
    C_{\alpha\beta\gamma\delta}={}^{(4)}R_{\alpha\beta\gamma\delta}+\frac{1}{2}\left(g_{\alpha\delta}{}^{(4)}R_{\beta\gamma}+g_{\beta\gamma}{}^{(4)}R_{\alpha\delta}-g_{\alpha\gamma}{}^{(4)}R_{\beta\delta}-g_{\beta\delta}{}^{(4)}R_{\alpha\gamma}\right)\\
    +\frac{1}{6}\left(g_{\alpha\gamma}g_{\beta\delta}-g_{\alpha\delta}g_{\beta\gamma}\right){}^{(4)}R, \\ 
\end{gathered} \label{conformal tensor}
\end{equation}
where ${}^{(4)}R_{\alpha\beta\gamma\delta}$ is the lowered Riemann tensor, ${}^{(4)}R_{\alpha\beta}$ is the Ricci tensor, and ${}^{(4)}R$ is the Ricci scalar. Weyl's tensor contains the ten field components that are not connected to the matter conditions of Einsteins tensor. The Weyl tensor has the symmetries $C_{\alpha\beta\gamma\delta}=-C_{\beta\alpha\gamma\delta}=C_{\gamma\delta\alpha\beta}=-C_{\alpha\beta\delta\gamma}$. We may use Gauss's equation to write the Weyl tensor's spatial components solely in terms of the canonical variables, $C_{abcd}={}^{(3)}R_{abcd}+K_{ac}K_{bd}-K_{ad}K_{bc}$.

We introduce the definition for the contraction of the normal vectors with the conformal metric tensor as
\begin{equation}
    n^{\delta}C_{abc\delta}~{:=}~C_{abc{L}}. \label{contract}\\
\end{equation}
We can construct D-invariant components of covariant tensors of arbitrary rank in this form. To do so, we start with a given covariant vector $v_{\mu}$. We then calculate the variation of $v_{\mu}$ under the infinitesimal diffeomorphism \eqref{diffeo} as, 
\begin{equation}
    \begin{aligned}
        \Bar{\delta}v_{\mu}=&~\epsilon^{\nu}_{,\mu}v_{\nu}+v_{\mu,\nu}\epsilon^{\nu}, \\
        \Bar{\delta}N=&~N\dot{\epsilon}^0+\dots,\\
        \Bar{\delta}N^a=&~N^a\dot{\epsilon}^0+\dot{\epsilon}^a+\dots, \notag
    \end{aligned}
\end{equation}
so 
\begin{equation}
\begin{aligned}
    \Bar{\delta}n^0=&~\Bar{\delta}\left(N^{-1}\right)\\
    =&-N^{-2}\Bar{\delta}N \\
    =&-N^{-1}\dot{\epsilon}^0+\dots, \notag
\end{aligned}
\end{equation}
and
\begin{equation}
\begin{aligned}
    \Bar{\delta}n^a=&-\Bar{\delta}\left(N^{-1}N^a\right) \\
    =&~N^{-2}\Bar{\delta}NN^a-N^{-1}\Bar{\delta}N^a \\
    =&~N^{-1}N^a\dot{\epsilon}^0-N^{-1}\left(N^a\dot{\epsilon}^0+\dot{\epsilon}^a\right)+\dots \\
    =&~N^{-1}\dot{\epsilon}^a+\dots. \notag
\end{aligned}
\end{equation}
Keeping only the time derivatives of $\epsilon^{\mu}$ we find,
\begin{equation}
\begin{aligned}
    \Bar{\delta}\left(v_{\mu}n^{\mu}\right)=&\left(\epsilon^{\nu}_{,\mu}v_{\nu}+v_{\mu,\nu}\epsilon^{\nu}\right)n^{\mu}+v_0\Bar{\delta}n^0-v_a\Bar{\delta}n^a\\
    =&~\dot{\epsilon}^{\nu}v_{\nu}N^{-1}-v_0N^{-1}\dot{\epsilon}^0-v_aN^{-1}\dot{\epsilon}^a+\dots. \notag
\end{aligned}
\end{equation}
This shows that the variation does not depend on time derivatives of $\epsilon^{\mu}$. Therefore, if we consider variations under diffeomorphisms that vanish on the constant $x_0$ hypersurface, $v_L:=v_{\mu}n^{\mu}$ is a D-invariant \cite{bergmann_1989,salisbury_2020}. We should also note that for multiple contractions the tensor is denoted similarly as
\begin{equation}
    n^{\beta}n^{\delta}C_{a\beta{c}\delta}~{:=}~n^{\beta}C_{a\beta{cL}}~{:=}~C_{a{L}c{L}}=C_{abcd}e^{bd}. \label{dblcontract}
\end{equation}
Multiple normal vector contractions, $L$, are organized so that they are always the last index for the conformal tensor's three-forms, and the even indices for the conformal tensor's two-forms.

\subsection{Tetrad Formalism}

We must now create an orthonormal tetrad composed of the vector normal to the time-constant hypersurfaces $n^{\mu}$ and triad vectors. We choose the tetrad so that the terms $E^{\mu}_{I}$ are the orthogonal unit vectors. In particular we take 
\begin{subequations}
    \begin{equation}
        E^{\mu}_{0}=n^{\mu}=\delta^{\mu}_0N^{-1}-\delta^{\mu}_aN^{-1}N^a, \label{tetrad 0}
    \end{equation}
    \begin{equation}
        E^{\mu}_i=\delta^{\mu}_a\delta^{a}_i \label{tetrad i}.
    \end{equation}
\end{subequations}
Then the full contravariant tetrad set is written as
\begin{equation}
\begin{gathered}
    E^{\mu}_{I}=
        \begin{pmatrix} 
         N^{-1}  & 0 \\
            -N^{-1}N^a & T^a_{i} \\
        \end{pmatrix}, ~~
\end{gathered}
\end{equation}
where we refer to the $T^{a}_i$ as the contravariant triads. This triad is related to the inverse spatial metric by $e^{ab}=T^{a}_iT^{b}_i$.
The corresponding covariant tetrad is written in terms of covariant triads represented by the orthonormal spatial one-forms $t^i_a$, where our 3-metric is given by $g_{ab}=t^i_at^i_b$ \cite{pons_gauge_reality_2000}. The full covariant tetrad is then given by
\begin{equation}
\begin{gathered}
    e^I_{\mu}=
        \begin{pmatrix} 
         N  & 0 \\
            t^i_aN^a & t^i_a \\
        \end{pmatrix}.
\end{gathered}
\end{equation}

\subsection{The Enlarged Bergmann-Komar Group}

In this section we review content of \cite{bergmann_1962} in a more modern, readable manner. G\'eh\'eniau and Debever first showed in \cite{geheniau_1956} that the Weyl tensor gives four algebraically indepentent scalars. This approach utilizes the fact that in the absence of any gravitational sources, the Weyl and Riemann tensors are the same. Bergmann reformulated this slightly using six linearly independent bivectors (skewsymmetric tensors of rank 2) that represent the Weyl tensor in phase-space (following the bivector decomposition of Petrov \cite{petrov2000classification}). The index pairs are:
\begin{equation}
\begin{gathered}
       \begin{matrix} 
         \text{Bivector index}~(\mathcal{A,B,\dots})  & \vline & \text{Index pair} \\
         \hline
            1 & \vline & (23) \\
            2 & \vline & (31) \\
            3 & \vline & (12) \\
            4 & \vline & (10) \\
            5 & \vline & (20) \\
            6 & \vline & (30) \\
        \end{matrix} ~~
        \\
\end{gathered}
\end{equation}
Note that for moving between the bivector and metric form, the index relations used for calculations are
\begin{equation}
\begin{gathered}
    \mathcal{A}\longleftrightarrow\alpha\beta,~~\mathcal{B}\longleftrightarrow\gamma\delta,\\
    \mathcal{C}\longleftrightarrow\rho\sigma,~~\mathcal{D}\longleftrightarrow\mu\nu,\\
    \mathcal{E}\longleftrightarrow\kappa\lambda,~~\mathcal{F}\longleftrightarrow\tau\omega. \label{indices} \\ 
\end{gathered}
\end{equation}
The Weyl tensor can then be represented as a $6\times{6}$ matrix $C$, where we must define two additional matrices. One will take on the role of our metric tensor, $G$, and the other will raise and lower bivector indices
\begin{subequations}
    \begin{equation}
        G^{\mathcal{BC}}:=G^{[\gamma\delta][\rho\sigma]}=g^{\gamma\rho}g^{\delta\sigma}-g^{\delta\rho}g^{\gamma\sigma}, \label{G}\\
    \end{equation}
    \begin{equation}
        {\aleph^\mathcal{A}}_\mathcal{B}={\aleph^{\alpha\beta}}_{\gamma\delta}=\frac{1}{\sqrt{-^{(4)}g}}g_{\rho\gamma}g_{\sigma\delta}\epsilon^{\alpha\beta\rho\sigma}. \label{aleph}\\
    \end{equation}
\end{subequations}
Here ${\epsilon}^{\alpha\beta\rho\sigma}$ is the Levi-Civita tensor density, with ${\epsilon}^{0123}=1$. To avoid confusion with regard to Bergmann's notation, we will use epsilon to denote the Levi-Civita tensor with coordinate indices and the delta symbol to represent Minkowski indices. Note that this epsilon has a tensor density of weight one, e.g.
\begin{equation}
    {\epsilon}'^{0123}={\epsilon}^{\alpha\beta\rho\sigma}\frac{{\partial}x'^0}{{\partial}x^{\alpha}}\frac{{\partial}x'^1}{{\partial}x^{\beta}}\frac{{\partial}x'^2}{{\partial}x^{\rho}}\frac{{\partial}x'^3}{{\partial}x^{\sigma}}\left|\frac{{\partial}x}{{\partial}x'}\right|=1. \\
\end{equation}
Additionally, $G$ and $\aleph$ satisfy the relations
\begin{subequations}
\begin{equation}
    \aleph^2=-\mathbb{I}_6, 
\end{equation}
\begin{equation}
    \aleph{G}=\left(\aleph{G}\right)^{T}=-G\aleph^{{T}}, 
\end{equation}
\end{subequations}
where the superscript ${T}$ indicates transpose, and $\mathbb{I}_6$ denotes the $6\times6$ unit matrix. We can then raise the indices of $\aleph$ by
\begin{equation}
\begin{gathered}
    \aleph^{\mathcal{DA}}={{\aleph}^{\mathcal{D}}}_{\mathcal{E}}G^{\mathcal{EA}}. \label{bivectorraise} \\
\end{gathered}
\end{equation}
The algebraic properties of the Weyl matrix $C$ are
\begin{subequations}
\begin{equation}
    C=C^T, \\
\end{equation}
\begin{equation}
     C\aleph={{}^*C}={{}^*C^{T}}={\aleph^{T}C},\\
\end{equation}
\begin{equation}
    \text{Tr}(C)=0, \\
\end{equation}
\begin{equation}
    \text{Tr}({{}^*C})=0. \\
\end{equation}
\end{subequations}
We transform the above algebraic relations using tetrads of mutually orthogonal unit-vectors at a {``world point''} and distinguish the first three bivector indices from the second triplet by denoting
\begin{equation}
\begin{gathered}
    G=
        \begin{pmatrix} 
         \mathbb{I}_3  & 0 \\
            0 & -\mathbb{I}_3 \\
        \end{pmatrix}, ~~
    \aleph=
        \begin{pmatrix} 
         0  & -\mathbb{I}_3 \\
            \mathbb{I}_3 & 0 \\
        \end{pmatrix}.
\end{gathered}
\end{equation}
We may write the algebraic properties of matrix $C$ as a set of $3\times3$ matrices $A$ and $B$,
\begin{subequations}
\begin{equation}
    C=
        \begin{pmatrix} 
         A  & B \\
            B & -A \\
        \end{pmatrix}, ~~
        \\
\end{equation}
\begin{equation}
     A=A^{T},\\
\end{equation}
\begin{equation}
    B=B^{T}, \\
\end{equation}
\begin{equation}
    \text{Tr}(A)=0, \\
\end{equation}
\begin{equation}
    \text{Tr}(B)=0. \\
\end{equation}
\end{subequations}
After a Lorentz transformation, $C$ may be written in complex notation as a symmetric, complex, and trace-free matrix
\begin{equation}
    \boldsymbol{Q}=A+iB, \\
\end{equation}
which transforms as a tensor rank-2, and is diagonalized via Petrov classification \cite{zakhary_2003,stephani_exact_2009}.

The Weyl scalars may be derived by casting the Weyl tensor into an eigenvalue problem in which the metric tensor becomes Minkowskian when the two symmetric 3D matrices, $A$ and $B$, traces vanish \textit{individually}
\begin{equation}
    \left(C_{\alpha\beta\gamma\delta}-\lambda\left(g_{\alpha\gamma}g_{\beta\delta}-g_{\alpha\delta}g_{\beta\gamma}\right)\right)V^{\gamma\delta}=0, \\
\end{equation}
where the skewsymmetric tensor $V^{\gamma\delta}$ is the eigen-bivector and $\lambda$ an eigenvalue \cite{pirani_1957,bk_1960}. The solution of this problem gives four scalars:
\begin{subequations}
\begin{equation}
    W^1:=\text{Tr}\left(CGCG\right)=C_{\mathcal{AB}}G^{\mathcal{BC}}C_{\mathcal{CD}}G^{\mathcal{DA}}, \label{W1}
\end{equation}
\begin{equation}
    W^2:=\text{Tr}\left(CGC\aleph\right)=C_{\mathcal{AB}}G^{\mathcal{BC}}C_{\mathcal{CD}}\aleph^{\mathcal{DA}}, \label{W2}
\end{equation}
\begin{equation}
    W^3:=\text{Tr}\left(CGCGCG\right)=C_{\mathcal{AB}}G^{\mathcal{BC}}C_{\mathcal{CD}}G^{\mathcal{DE}}C_{\mathcal{EF}}G^{\mathcal{FA}}, \label{W3}\\
\end{equation}
\begin{equation}
    W^4:=\text{Tr}\left(CGCGC\aleph\right)=C_{\mathcal{AB}}G^{\mathcal{BC}}C_{\mathcal{CD}}G^{\mathcal{DE}}C_{\mathcal{EF}}\aleph^{\mathcal{FA}}. \label{W4} \\
\end{equation}
\end{subequations}
Here $W^1, W^2$ are quadratic, and $W^3, W^4$ are cubic expressions of the Weyl conformal tensor where the scalars are written in terms of canonical variables \cite{bk_1960}\footnote{Bergmann originally wrote the scalars as ``$A^I$".}. 

For a set of four independent functions of the scalars to serve as an ICS (or ICRF) they must be algebraically independent, noting that not all solutions to Einstein's field equations obey this requirement. This is evident for Petrov types, $N$ and $III$, where this set of coordinate-invariants vanish, although the Weyl tensor itself is nonzero \cite{sibgatullin_1991}.

Since $W^2$ ($W^4$) requires the use of both the tensor $\aleph^{\mathcal{DA}}$ ($\aleph^{\mathcal{FA}}$) and the triads used to describe this matrix, we derive $\aleph^{\mathcal{DA}}$ below to ensure invariance
We wish to rewrite the expression \eqref{aleph} in terms of the Minkowski Levi-Civita symbol $\delta^{IJKM}$, where $\delta^{0123}=1$. We note that there must exist a constant $\mathfrak{K}$ satisfying
\begin{equation}
    {\epsilon}^{\alpha\beta\rho\sigma}={\mathfrak{K}}\delta^{IJKM}E^{\alpha}_{I}E^{\beta}_{J}E^{\rho}_{K}E^{\sigma}_{M}. \label{lc_tensor_density}
\end{equation}
To determine the constant $\mathfrak{K}$, we contract with the covariant tetrads to get 
\begin{equation}
\begin{gathered}
    \epsilon^{\alpha\beta\rho\sigma}e^I_{\alpha}e^J_{\beta}e^K_{\rho}e^M_{\sigma}={\mathfrak{K}}\delta^{IJKM}, \\
\end{gathered}
\end{equation}
so that
\begin{equation}
\begin{gathered}
    {\mathfrak{K}}=\epsilon^{\alpha\beta\rho\sigma}e^0_{\alpha}e^1_{\beta}e^2_{\rho}e^3_{\sigma}=N\delta^{abc}t^1_at^2_bt^3_c=Nt.
\end{gathered}
\end{equation}
We then can write \eqref{lc_tensor_density} in the form
\begin{equation}
    \epsilon^{\alpha\beta\rho\sigma}=Nt{\delta}^{IJKM}E^{\alpha}_IE^{\beta}_JE^{\rho}_KE^{\sigma}_M. \label{lc2}
\end{equation}
Note that
\begin{equation}
    \sqrt{-^{(4)}g}={Nt}, \label{Nt}
\end{equation}
so that \eqref{lc2} becomes
\begin{equation}
    \epsilon^{\alpha\beta\rho\sigma}={\delta}^{IJKM}E^{\alpha}_IE^{\beta}_JE^{\rho}_KE^{\sigma}_M\sqrt{-^{(4)}g}. \label{lc3}
\end{equation}
Substituting the above equation into \eqref{aleph} gives
\begin{equation}
\begin{gathered}
    {\aleph^\mathcal{A}}_\mathcal{B}={\delta}^{IJKM}E^{\alpha}_IE^{\beta}_JE^{\rho}_KE^{\sigma}_Mg_{\rho\gamma}g_{\sigma\delta}.
\end{gathered}
\end{equation}
Re-indexing $\mathcal{A}\rightarrow{\mathcal{D}}$, and $\mathcal{B}\rightarrow{\mathcal{E}}$ we can write
\begin{equation}
\begin{gathered}
    {\aleph^\mathcal{D}}_\mathcal{E}={\aleph^{\mu\nu}}_{\kappa\lambda}=g_{\alpha\kappa}g_{\beta\lambda}\delta^{IJKM}E^{\mu}_IE^{\nu}_JE^{\alpha}_KE^{\beta}_M.
\end{gathered}
\end{equation}
Raising with \eqref{bivectorraise} gives
\begin{equation}
\begin{gathered}
    {\aleph^{\mathcal{DA}}}=\frac{1}{2}g_{\alpha\kappa}g_{\beta\lambda}\delta^{IJKM}E^{\mu}_IE^{\nu}_JE^{\alpha}_KE^{\beta}_MG^{\mathcal{EA}}. \notag
\end{gathered}
\end{equation}
Using \eqref{G} for $G^{\mathcal{EA}}$ and ${\delta_{\alpha}}^{\alpha}=g_{\alpha\kappa}g^{\alpha\kappa}$, the above simplifies to 
\begin{equation}
\begin{gathered}
    {\aleph^{\mathcal{DA}}}=\delta^{IJKM}E^{\mu}_IE^{\nu}_JE^{\alpha}_KE^{\beta}_M.  \notag
\end{gathered}
\end{equation}
Rewriting in terms of \eqref{unit_normal} gives us
\begin{equation}
\begin{gathered}
    {\aleph^{\mathcal{DA}}}
    =\delta^{ijk}(n^{\mu}{{E_i}^{\nu}}{{E_j}^{\alpha}}{{E_k}^{\beta}}-n^{\nu}{{E_i}^{\mu}}{{E_j}^{\alpha}}{{E_k}^{\beta}}+{{E_i}^{\mu}}{{E_j}^{\nu}}n^{\alpha}{{E_k}^{\beta}}-{{E_i}^{\mu}}{{E_j}^{\nu}}{{E_k}^{\alpha}}n^{\beta}).  \label{aleph DA}
\end{gathered}
\end{equation}
We now show that the scalar densities that require $\aleph$ transform properly under arbitrary spatial diffeomorphisms:
\begin{equation}
\begin{aligned}
    T(x)&=\delta^{ijk}T_i^aT_j^bT_k^c, \\
    T'(x')&=\delta^{ijk}T_i^dT_j^eT_k^f\frac{{\partial}{x'^a}}{{\partial}{x^d}}\frac{{\partial}{x'^b}}{{\partial}{x^e}}\frac{{\partial}{x'^c}}{{\partial}{x^f}}, \\
    {\epsilon}^{def}&=t\delta^{ijk}T_i^dT_j^eT_k^f=T^{-1}\delta^{ijk}T_i^dT_j^eT_k^f, \\
     T'(x')&=T{\epsilon}^{def}\frac{{\partial}{x'^a}}{{\partial}{x^d}}\frac{{\partial}{x'^b}}{{\partial}{x^e}}\frac{{\partial}{x'^c}}{{\partial}{x^f}}=T\left|\frac{{\partial}x'}{{\partial}x}\right|,  \notag
\end{aligned}
\end{equation}
(so the determinant transforms as a scalar density of weight $-1$) while $\epsilon^{abc}$ transforms as a density of weight $+1$. Therefore we require this additional factor $T$ so that, $W^2$ (or $W^4$) transforms as a scalar (or scalar density of weight zero):
\begin{equation}
         \therefore W^2(x')=W^2(x)\left|\frac{{\partial}x'}{{\partial}x}\right|\left|\frac{{\partial}x}{{\partial}x'}\right|=W^2(x). \notag
\end{equation}

Since both $W_2$ and $W_4$ require the scalar density $T(x)$, the presence of $T(x)$ in $W_4$ ensures that $W_4$ also transforms as a scalar of weight zero under the spatial diffeomorphisms. The only difference is that $W_2$ couples linearly to $\aleph^{\mathcal{DA}}$ where $W_4$ couples by a cubic contraction (of $\aleph^{\mathcal{FA}}$).

\section{Solution of Weyl Scalars}

We now derive the exact relations between the four Weyl scalar functional invariants, $W^I$, and the canonical phase space variables. This was proposed in 1962 by Peter Bergmann after two were derived by Bergmann and Komar two years prior \cite{bk_1960,bergmann_1962}.

\subsection{The First Weyl Scalar}
The first Weyl scalar does not require the tensor $\aleph$ and so we shall work solely with the ADM formalism. Beginning with
\begin{equation}
\begin{gathered}
    W^1=C_{\mathcal{AB}}G^{\mathcal{BC}}C_{\mathcal{CD}}G^{\mathcal{DA}}, \tag{\ref{W1}}
\end{gathered}
\end{equation}
we rewrite the first Weyl scalar as
\begin{equation}
    W^{1}=C_{\alpha\beta\gamma\delta}C_{\rho\sigma\mu\nu}\left(G^{\mathcal{BC}}G^{\mathcal{DA}}\right). \tag{32a$'$} \label{32a$'$}
\end{equation}
Using \eqref{WG1}, we find 
\begin{equation}
\begin{gathered}
    W^{1}=C_{\alpha\beta\gamma\delta}C_{\rho\sigma\mu\nu}\left(4e^{\gamma\rho}e^{\delta\sigma}e^{\mu\alpha}e^{\nu\beta}-16e^{\gamma\rho}e^{\delta\sigma}e^{\mu\alpha}n^{\nu}n^{\beta}+16e^{\gamma\rho}e^{\mu\alpha}n^{\delta}n^{\sigma}n^{\nu}n^{\beta}\right). \notag
\end{gathered}
\end{equation}
We contract using equations \eqref{contract} and \eqref{dblcontract}
\begin{equation}
\begin{gathered}
    W^{1}=4C_{\alpha\beta\gamma\delta}C_{\rho\sigma\mu\nu}e^{\gamma\rho}e^{\delta\sigma}e^{\mu\alpha}e^{\nu\beta}-16C_{\alpha{L}\gamma\delta}C_{\rho\sigma\mu{L}}e^{\gamma\rho}e^{\delta\sigma}e^{\mu\alpha}+16C_{\alpha{L}\gamma{L}}C_{\rho{L}\mu{L}}e^{\gamma\rho}e^{\mu\alpha}. \notag
\end{gathered}
\end{equation}
We simplify by using \eqref{DI1}
\begin{equation}
\begin{gathered}
    W^{1}=4C_{abcd}C_{efgh}e^{ce}e^{df}e^{ga}e^{hb}-16C_{a{L}cd}C_{efg{L}}e^{ce}e^{df}e^{ga}+16C_{a{L}c{L}}C_{e{L}g{L}}e^{ce}e^{ga}, \notag
\end{gathered}
\end{equation}
and the inverse spatial metric raises the conformal metric tensor indices, so 
\begin{equation}
\begin{gathered}
    W^{1}=4C_{abcd}C^{abcd}-16C_{a{L}cd}{C^{cda}}_{L}+ 16C_{a{L}c{L}}{{{{C^{c}}_{L}}^{a}}_{L}}. \notag
\end{gathered}
\end{equation}
The symmetries of the Weyl tensor allow us to re-index before (or after) raising the indices so that the above line becomes
\begin{equation}
    \begin{gathered}
        W^{1}=4C_{abcd}C^{abcd}-16C_{abcL}{C^{abc}}_{L}+16C_{a{L}b{L}}{{{{C^{a}}_{L}}^{b}}_{L}}. \notag
    \end{gathered}
\end{equation}
Simplifying the coefficients leaves the first Weyl scalar
\begin{subequations}
\begin{equation}
\begin{gathered}
    W^{1}=C_{abcd}C^{abcd}-4C_{abc{L}}{C^{abc}}_{L}+4C_{a{L}b{L}}{{{{C^{a}}_{L}}^{b}}_{L}}, 
\end{gathered}
\end{equation}
or written in the shorthand form so that we drop the $L$ indicies
\begin{equation}
\begin{gathered}
    \boxed{W^{1}=~C_{abcd}C^{abcd}-4C_{abc}{C^{abc}}+4C_{ab}{{C^{a}}^{b}}}. \label{W1final}
\end{gathered}
\end{equation}
\end{subequations}
This indicates that the original publication of Bergmann and Komar in \cite{bk_1960} is lacking a negative sign on the second term, $4C_{abc}{C^{abc}}$ in their equation $(3)$ of their first Weyl scalar $W^1$. This also fixes the missing negative sign reprinted in equation $(44)$ of the modern literature \cite{pons_2005}. This sign error may have arisen from the method of contracting for the on-shell (3- and 2-form) Weyl tensors. This is seen by Bergmann and Komar in \cite{bk_1960}, where they agree with our definition of \eqref{dblcontract} where Pons et al. \cite{pons_2005} found this term by disregarding the original negative sign when using their equation $(38)$. 

\subsection{The Second Weyl Scalar} 

Starting from the second Weyl scalar
\begin{equation}
\begin{gathered}
    W^2=C_{\mathcal{AB}}G^{\mathcal{BC}}C_{\mathcal{CD}}\aleph^{\mathcal{DA}}, \tag{\ref{W2}}
\end{gathered}
\end{equation}
we re-order, $W^2=G^{\mathcal{BC}}C_{\mathcal{CD}}\aleph^{\mathcal{DA}}C_{\mathcal{AB}}$, and since we are using $\aleph$ in our calculation we should turn our attention to calculating the product of
\begin{equation}
    \Upsilon:=C_{\mathcal{CD}}\aleph^{\mathcal{DA}}C_{\mathcal{AB}}. \label{alephproduct}
\end{equation}
First by substituting \eqref{aleph DA} into \eqref{alephproduct} and re-indexing we get
\begin{equation}
    \Upsilon=C_{\rho\sigma\mu\nu}\delta^{ijk}(n^{\mu}{{E_i}^{\nu}}{{E_j}^{\alpha}}{{E_k}^{\beta}}-n^{\nu}{{E_i}^{\mu}}{{E_j}^{\alpha}}{{E_k}^{\beta}}+{{E_i}^{\mu}}{{E_j}^{\nu}}n^{\alpha}{{E_k}^{\beta}}-{{E_i}^{\mu}}{{E_j}^{\nu}}{{E_k}^{\alpha}}n^{\beta})C_{\alpha\beta\gamma\delta}. \label{43$'$} \tag{43$'$}
\end{equation}
Contracting \eqref{43$'$} with \eqref{contract} becomes
\begin{equation}
\begin{gathered}
    \Upsilon=\delta^{ijk}(C_{\rho\sigma{L}\nu}{{E_i}^{\nu}}{{E_j}^{\alpha}}{{E_k}^{\beta}}C_{\alpha\beta\gamma\delta}-C_{\rho\sigma\mu{L}}{{E_i}^{\mu}}{{E_j}^{\alpha}}{{E_k}^{\beta}}C_{\alpha\beta\gamma\delta}\\
    +C_{\rho\sigma\mu\nu}{{E_i}^{\mu}}{{E_j}^{\nu}}{{E_k}^{\beta}}C_{{L}\beta\gamma\delta}-C_{\rho\sigma\mu\nu}{{E_i}^{\mu}}{{E_j}^{\nu}}{{E_k}^{\alpha}}C_{\alpha{L}\gamma\delta}).  \notag
\end{gathered}
\end{equation}
Re-indexing again shows that the terms are equivalent, yielding 
\begin{equation}
    \Upsilon=4C_{\rho\sigma{L}\nu}{{E_i}^{\nu}}{{E_j}^{\alpha}}{{E_k}^{\beta}}C_{\alpha\beta\gamma\delta}\delta^{ijk}.\\ \label{upsilon}
\end{equation}
Substituting \eqref{tetrad i} into \eqref{upsilon} returns
\begin{equation}
\begin{gathered}
    \Upsilon=4C_{\rho\sigma{L}\nu}{{T_i}^a}{{T_j}^b}{{T_k}^c}C_{\alpha\beta\gamma\delta}{\delta_a}^{\nu}{\delta_b}^{\alpha}{\delta_c}^{\beta}\delta^{ijk},  \notag
\end{gathered}
\end{equation}
and simplifying gives
\begin{equation}
\begin{gathered}
    \Upsilon=~-4\delta^{ijk}C_{\rho\sigma{aL}}{{T_i}^a}{{T_j}^b}{{T_k}^c}C_{bc\gamma\delta}.   \notag
\end{gathered}
\end{equation}
We can further simplify with the equality
\begin{equation}
    \delta^{ijk}T_i^aT_j^bT_k^c=T\epsilon^{abc}, 
\end{equation}
to give us \eqref{alephproduct} in the form\footnote{Note that the Greek indices in equation \eqref{alephproduct} are dependent only on the first index and last index.}
\begin{equation}
\begin{gathered}
    \Upsilon=C_{\mathcal{CD}}\aleph^{\mathcal{DA}}C_{\mathcal{AB}}=-4T\epsilon^{abc}C_{\rho\sigma{aL}}C_{bc\gamma\delta}.   \label{w2aleph}
\end{gathered}
\end{equation}
 With \eqref{w2aleph} and \eqref{WG2} we write \eqref{W2} as
\begin{equation}
    \begin{gathered}
        W^2=\left(2e^{\gamma\rho}e^{\delta\sigma}-2^2e^{\gamma\rho}n^{\delta}n^{\sigma}\right)\times\left(-4T\epsilon^{abc}C_{\rho\sigma{aL}}C_{bc\gamma\delta}\right). \label{32b$'$} \tag{32b$'$}
    \end{gathered}
\end{equation}
Contracting \eqref{32b$'$} with \eqref{contract} then simplifying with \eqref{DI1} gives 
\begin{equation}
    \begin{gathered}
        W^2=-4T\epsilon^{abc}\left(2C_{ef{aL}}C_{bcgd}e^{ge}e^{df}-2^2C_{eLaL}C_{bcgL}{e^{ge}}\right), \notag 
    \end{gathered}
\end{equation}
then raising with the inverse spatial metric we have
\begin{equation}
    \begin{gathered}
        W^2={-8T\epsilon^{abc}(C_{ef{aL}}{C_{bc}}^{ef}}-2{{C^{g}}_{{LaL}}C_{bcg{L}}}). \notag
    \end{gathered}
\end{equation}
Re-indexing we find \eqref{W2}
\begin{subequations}
    \begin{equation}
    \begin{gathered}
        W^2={-8T\epsilon^{abc}(C_{de{aL}}{C_{bc}}^{de}}-2{{C^{d}}_{{LaL}}C_{bcd{L}}}),
    \end{gathered}
\end{equation}
or, 
\begin{equation}
    \boxed{W^2=T\epsilon^{abc}\left({-C_{abc{L}}{C_{de}}^{ab}}+2{C_{abc{L}}}{C^{c}}_{{LdL}}\right)}. \label{W2final}
\end{equation}
\end{subequations}
We note that there is a freedom when raising the indicies and that the authors choose the least-mixed terms (e.g. mixed 2-form, lowered 3-form). We compare this to the original proposal from Bergmann and Komar in \cite{bk_1960} where they presented their solution of the second Weyl scalar as 
\begin{equation}
\begin{gathered}
    W^2=\epsilon^{lms}\left(C_{iklm}{C^{ik}}_{s}+2C_{klm}{C^k}_{s}\right). \notag
\end{gathered}
\end{equation}
We re-index and apply our method of normal unit contractions of $L$, as well as the symmetry \eqref{symmetry} for the sign of the first term
\begin{align*}
    W^2 &=\epsilon^{abc}\left(C_{deab}{C^{de}}_{c}+2C_{cab}{C^c}_{d}\right) \\
    &=\epsilon^{abc}\left(C_{deab}{C^{de}}_{cL}+2C_{cLab}{C^c}_{LdL}\right) \\
    &=\epsilon^{abc}\left(C_{deaL}{C^{de}}_{cb}+2C_{cLab}{C^c}_{LdL}\right) \\
    &=\epsilon^{abc}\left(-C_{deaL}{C_{bc}}^{de}+2C_{cLab}{C^c}_{LdL}\right) \\
    &=\epsilon^{abc}\left(-C_{abcL}{C_{de}}^{ab}+2C_{abcL}{C^c}_{LdL}\right). \notag
\end{align*}
So we are in agreement with the original publication by Bergmann and Komar in \cite{bk_1960} with the exception of the missing scalar-density $T$, as shown by Pons et. al \cite{pons_2005}.

\subsection{The Third Weyl Scalar}

Constructing the third Weyl scalar follows the method of the first
\begin{equation}
\begin{gathered}
    W^3=C_{\mathcal{AB}}G^{\mathcal{BC}}C_{\mathcal{CD}}G^{\mathcal{DE}}C_{\mathcal{EF}}G^{\mathcal{FA}}, \tag{\ref{W3}}\\
\end{gathered}
\end{equation}
however the direct proof grows exponentially and can be unwieldy. This is seen when solving for the product of $G^{\mathcal{BC}}G^{\mathcal{DE}}G^{\mathcal{FA}}$. Substituting the product \eqref{WG3} into \eqref{W3} we find
\begin{equation}
\begin{gathered}
    W^3=\left(2^3e^{\gamma\rho}e^{\delta\sigma}e^{\mu\kappa}e^{\nu\lambda}e^{\tau\alpha}e^{\omega\beta}-2^4(e^{\gamma\rho}e^{\delta\sigma}e^{\mu\kappa}e^{\nu\lambda}e^{\tau\alpha}n^{\omega}n^{\beta}+2e^{\gamma\rho}e^{\delta\sigma}e^{\mu\kappa}e^{\tau\alpha}e^{\omega\beta}n^{\nu}n^{\lambda})\right.\\
            \left.+2^5e^{\gamma\rho}e^{\mu\kappa}e^{\tau\alpha}e^{\omega\beta}n^{\delta}n^{\sigma}n^{\nu}n^{\lambda}+2^6e^{\gamma\rho}e^{\delta\sigma}e^{\mu\kappa}e^{\tau\alpha}n^{\nu}n^{\lambda}n^{\omega}n^{\beta}\right)\times(C_{\alpha\beta\gamma\delta}C_{\rho\sigma\mu\nu}C_{\kappa\lambda\tau\omega}). \notag
\end{gathered}
\end{equation}
Contracting with \eqref{contract},\eqref{dblcontract}, and \eqref{DI1} in one step returns
\begin{equation}
    \begin{gathered}
        W^3=2^3e^{ce}e^{df}e^{gi}e^{hj}e^{ka}e^{lb}
   C_{abcd}C_{efgh}C_{ijkl}\\
        -2^4(e^{ce}e^{df}e^{gi}e^{hj}e^{ka}C_{aLcd}C_{efgh}C_{ijkL}+2e^{ce}e^{df}e^{gi}e^{ka}e^{lb}C_{abcd}C_{efgL}C_{iLkl})\\
            +2^5e^{ce}e^{gi}e^{ka}e^{lb}C_{abcL}C_{eLgL}C_{iLkl}
            +2^6e^{ce}e^{df}e^{gi}e^{ka}C_{aLcd}C_{efgL}C_{iLkL}. \notag
    \end{gathered}
\end{equation}
Raising\footnote{We raise so that the leading term matches with $w_2:= C_{abcd}{C^{cd}}_{ef}{C}^{efab}$ for the Carminati-McLenaghan invariants \cite{carminati1991algebraic}.} and re-indexing gives
\begin{equation}
    \begin{gathered}
        W^3=8C_{abcd}{C^{cd}}_{ef}{C}^{efab}-16C_{abcd}{{C^{abe}}_{L}}{{C^{cd}}_{eL}}
    -32C_{abcd}{{C^{abe}}_{L}}{{C^{cd}}_{eL}}\\
    +96C_{abc{L}}{{C^{ab}}_{dL}}{{{{C^{c}}_{L}}^{d}}_{L}}-64C_{a{L}b{L}}{{C^{b}}_{{L}c{L}}}{{{{C^{a}}_{L}}^{c}}_{L}}, \notag
    \end{gathered}
\end{equation}
then simplifying gives the third scalar:
\begin{subequations}
\begin{equation}
\begin{gathered}
    W^3=C_{abcd}{C^{cd}}_{ef}{C}^{efab}
    -6C_{abcd}{C^{abe}}_{{L}}{C^{cd}}_{{eL}} \\
    +12C_{abc{L}}{{C^{ab}}_{dL}}{{{{C^{c}}_{L}}^{d}}_{L}}
    -8C_{a{L}b{L}}{{C^{b}}_{{L}c{L}}}{{{{C^{a}}_{L}}^{c}}_{L}}, 
\end{gathered}
\end{equation}
or,
\begin{equation}
\begin{gathered}
    \boxed{W^3=C_{abcd}{C^{cd}}_{ef}{C}^{efab}
    -6C_{abcd}{C^{abe}}{C^{cd}}_{{e}}
    +12C_{abc}{{C^{ab}}_{d}}{{{{C^{cd}}}}}
    -8C_{ab}{{C^{b}}_{c}}{{{{C^{ac}}}}}}. \label{W3final}
\end{gathered}
\end{equation}
\end{subequations}

\subsection{The Fourth Weyl Scalar}
Following the procedure of our solution of $W^2$, we use $\aleph^{\mathcal{FA}}$ to solve for $W^4$
\begin{equation}
\begin{gathered}
    W^4=C_{\mathcal{AB}}G^{\mathcal{BC}}C_{\mathcal{CD}}G^{\mathcal{DE}}C_{\mathcal{EF}}\aleph^{\mathcal{FA}}. \tag{\ref{W4}}
\end{gathered}
\end{equation}
We write the product of $C_{\mathcal{EF}}\aleph^{\mathcal{FA}}C_{\mathcal{AB}}$, where we recall from earlier that the product depends only on the first and last indices so that $\aleph^{\mathcal{FA}}$ is
\begin{equation}
\begin{gathered}
    \aleph^{\mathcal{FA}}=\delta^{ijk}(n^{\mu}{{E_i}^{\nu}}{{E_j}^{\tau}}{{E_k}^{\omega}}-n^{\nu}{{E_i}^{\mu}}{{E_j}^{\tau}}{{E_k}^{\omega}}+{{E_i}^{\mu}}{{E_j}^{\nu}}n^{\tau}{{E_k}^{\omega}}-{{E_i}^{\mu}}{{E_j}^{\nu}}{{E_k}^{\tau}}n^{\omega}). 
\end{gathered}
\end{equation}
Operating with the outer terms $C_{\mathcal{EF}}$ and $C_{\mathcal{AB}}$, produces
\begin{equation}
\begin{gathered}
    C_{\mathcal{EF}}\aleph^{\mathcal{FA}}C_{\mathcal{AB}}={-4T\epsilon^{abc}C_{\kappa\lambda{aL}}C_{bc\gamma\delta}}.  \label{W4aleph}
\end{gathered}
\end{equation}
Using the above \eqref{W4aleph} and substituting $G^{\mathcal{BC}}G^{\mathcal{DE}}$ \eqref{WG4}, $W^4$ becomes
\begin{equation}
    \begin{gathered}
        W^4=\left(-2^2e^{\gamma\rho}e^{\delta\sigma}e^{\mu\kappa}e^{\nu\lambda}-2^4\left(e^{\gamma\rho}e^{\delta\sigma}e^{\mu\kappa}n^{\nu}n^{\lambda}+e^{\gamma\rho}e^{\mu\kappa}e^{\nu\lambda}n^{\delta}n^{\sigma}\right)\right.\\
        \left.-2^5e^{\gamma\rho}e^{\mu\kappa}n^{\delta}n^{\sigma}n^{\nu}n^{\lambda}\right)\times\left({-4T\epsilon^{abc}{C_{\rho\sigma\mu\nu}}C_{\kappa\lambda{aL}}C_{bc\gamma\delta}}\right). \notag
    \end{gathered}
\end{equation}
Contracting with \eqref{contract} and \eqref{dblcontract} then using \eqref{DI1} to drop to spatial indicies,\footnote{We distinguish $a,b,c:=a',b',c'$ to ensure proper ordering of the contracted indices.} we find
\begin{equation}
    \begin{aligned}
        W^4= &T\epsilon^{abc}\left(-32e^{ce}e^{df}e^{gi}e^{hj}C_{efgh}C_{ij{a'L}}C_{b'c'cd}\right.\\
        &-2^4\left(e^{ce}e^{df}e^{gi}C_{efgL}C_{iL{a'L}}C_{b'c'cd}+e^{ce}e^{gi}e^{hj}C_{eLgh}C_{ij{a'L}}C_{b'c'cL}\right) \\
        &\left.-2^5e^{ce}e^{gi}C_{eLgL}C_{iL{a'L}}C_{b'c'cL}\right). \notag
    \end{aligned}
\end{equation}
Raising indicies with the inverse spatial metric, then re-indexing returns
\begin{equation}
    \begin{gathered}
        W^4=T\epsilon^{abc}\left(-32C_{fg{aL}}C_{bcde}C^{defg}+64({{{C^{def}}_L}}C_{f{L}{aL}}C_{bcde}+{{C^{efd}}_L}C_{bcd{L}}C_{ef{aL}})\right.\\
        \left.-128{{{{C^d}_L}^e}_L}C_{e{L}{aL}}C_{bcd{L}}\right). \notag 
    \end{gathered}
\end{equation}
Re-indexing again and simplifying, we find the fourth Weyl scalar:
\begin{subequations}
    \begin{equation}
    \begin{gathered}
        W^4=T\epsilon^{abc}\left(-C_{efg{L}}C_{abcd}C^{cdef}+2({{{C^{cde}}_L}}C_{e{L}{fL}}C_{abcd}+{{C^{dec}}_L}C_{abc{L}}C_{def{L}})\right.\\
        \left.-4{{{{C^c}_L}^d}_L}C_{d{L}{eL}}C_{abc{L}}\right),
    \end{gathered}
\end{equation}
or,
\begin{equation}
\begin{gathered}
     \boxed{W^{4}=T\epsilon^{abc}\left(-C_{abcd}C^{cdef}C_{efg}+2C_{abcd}C^{cde}C_{ef}+2C_{abc}C_{def}C^{dec}-4C_{abc}C^{cd}C_{de}\right)}. \label{W4final}
\end{gathered}
\end{equation}
\end{subequations}

\subsection{Weyl scalars in Tensorial Form}

We now present the set of all four Weyl scalars in terms of canonical phase-space variables:
\begin{subequations}
\begin{equation}
\begin{gathered}
    \boxed{W^{1}=~C_{abcd}C^{abcd}-4C_{abc}{C^{abc}}+4C_{ab}{{C^{a}}^{b}},} \tag{\ref{W1final}}
\end{gathered}
\end{equation}
\begin{equation}
\begin{gathered}
    \boxed{W^2=T\epsilon^{abc}\left({-C_{abc}{C_{de}}^{ab}}+2{{C^{c}}_{{a}}C_{abc}}\right),} \tag{\ref{W2final}}
\end{gathered}
\end{equation}
\end{subequations}
\begin{equation}
\begin{gathered}
    \boxed{W^3=C_{abcd}{C^{cd}}_{ef}{C}^{efab}
    -6C_{abcd}{C^{abe}}{C^{cd}}_{{e}} 
    +12C_{abc{}}{{C^{ab}}_{d}}{{{{C^{cd}}}}}
    -8C_{ab}{{C^{b}}_{c}}{{{{C^{ac}}}}},} \tag{\ref{W3final}}
\end{gathered}
\end{equation}
\begin{equation}
\begin{gathered}
    \boxed{W^{4}=T\epsilon^{abc}\left(-C_{abcd}C^{cdef}C_{efg}+2C_{abcd}C^{cde}C_{ef}+2C_{abc}C_{def}C^{dec}-4C_{abc}C^{cd}C_{de}\right).} \tag{\ref{W4final}}
\end{gathered}
\end{equation}

\section{Weyl Scalars for the Schwarzschild Solution}
We present the example for the Schwarzschild solution as by choosing our coordinates/ gauge we freeze our dynamical interpretation. Starting with $n^{\mu}$, the normal to the constant time hypersurface: 
\begin{equation}
    n^{\mu}=-g^{0\mu}\left(-g^{00}\right)^{-1/2}=\left(N^{-1},-N^{-1}N^a\right), \label{unit_normal} \\
\end{equation}
where
\begin{equation}
    n^{\mu}n_{\mu}=-1, \\
\end{equation}
so that
\begin{equation}
    n_{\mu}=\{-N,0,0,0\}, \\
\end{equation}
and
\begin{equation}
\begin{gathered}
    g^{00}=-\frac{1}{N^2}, \quad
    N^2=-\frac{1}{g^{00}}, \quad
    N=\frac{1}{\sqrt{-g^{00}}}. \\
\end{gathered}
\end{equation}
The lowered and raised normal vectors are then
\begin{equation}
\begin{gathered}
    n^{\mu}=\left\{\frac{1}{\sqrt{\left(1-\frac{2m}{r}\right)}},0,0,0\right\}, \quad
    n_{\mu}=\left\{-\sqrt{1-\frac{2m}{r}},0,0,0\right\}.
\end{gathered}
\end{equation}
For the Schwarzschild solution the contravariant metric is written
\begin{equation}
\begin{gathered}
    g^{\alpha\beta}=
        \begin{pmatrix} 
         -\frac{r}{r-2m}  & 0 & 0 & 0\\
            0 & 1-\frac{2m}{r} & 0 & 0\\
            0 & 0 & \frac{1}{r^2} & 0\\
            0 & 0 & 0 & \frac{1}{r^2\sin^2{\theta}}\\
        \end{pmatrix},
\end{gathered}
\end{equation}
so that
\begin{equation}
\begin{gathered}
    e^{\alpha\beta}=
        \begin{pmatrix} 
         0  & 0 & 0 & 0\\
            0 & 1-\frac{2m}{r} & 0 & 0\\
            0 & 0 & \frac{1}{r^2} & 0\\
            0 & 0 & 0 & \frac{1}{r^2\sin^2{\theta}}\\
        \end{pmatrix}, \quad\quad 
    e^{ab}=
        \begin{pmatrix} 
        1-\frac{2m}{r} & 0 & 0\\
        0 & \frac{1}{r^2} & 0\\
        0 & 0 & \frac{1}{r^2\sin^2{\theta}}\\
        \end{pmatrix}. 
\end{gathered}
\end{equation}
Or simply listed:
\begin{equation}
\begin{gathered}
    e^{11}=1-\frac{2m}{r}, \quad
    e^{22}=\frac{1}{r^2}, \quad
    e^{33}=\frac{1}{r^2\sin^2{\theta}}. \\ \label{spatialterms}
\end{gathered}
\end{equation}
The elements of the lowered Weyl tensor are the same as those of the lowered Riemann tensor:
\begin{equation}
\begin{aligned}
    C_{0101}=-\frac{2m}{r^3},& \quad
    C_{0202}=\frac{m(r-2m)}{r^2},& 
    &C_{0303}=\frac{(r-2m)m \sin^2{\theta}}{r^2}, \\
    C_{1212}=-\frac{m}{r-2m},& \quad
    C_{1313}=-\frac{m \sin^2{\theta}}{r-2m},& 
    &C_{2323}={2 m r\sin^2{\theta}}.  \label{loweredweylterms}
\end{aligned}
\end{equation}
To show that $C_{\alpha\beta}:=n^{\gamma}n^{\delta}C_{\alpha\gamma\beta\delta}:=C_{\alpha\gamma\beta\delta}e^{\gamma\delta}$ we evaluate 
\begin{equation}
\begin{gathered}
    C_{11}=C_{1212}e^{22}+C_{1313}e^{33}, \quad
    C_{22}=C_{2121}e^{11}+C_{2323}e^{33}, \quad 
    C_{33}=C_{3131}e^{11}+C_{3232}e^{22},  \label{2formlowered}
\end{gathered}
\end{equation}
substituting \eqref{spatialterms} and \eqref{loweredweylterms} into \eqref{2formlowered} we get 
\begin{equation}
\begin{gathered}
    C_{ab}=
        \begin{pmatrix} 
        -\frac{2m}{r^2\left(r-2m\right)} & 0 & 0\\
        0 & \frac{m}{r} & 0\\
        0 & 0 & \frac{{m \sin^2{\theta}}}{r}\\
        \end{pmatrix}.
\end{gathered}
\end{equation}
Alternatively, finding $C_{\alpha\beta\gamma}=-n^{\mu}C_{\mu\alpha\beta\gamma}$,
\begin{equation}
    n^0=\frac{1}{\sqrt{1-\frac{2m}{r}}},
\end{equation}
\begin{equation}
\begin{aligned}
    C_{101}&=n^0C_{0101}=\frac{1}{\sqrt{1-\frac{2m}{r}}}\left(-\frac{2m}{r^3}\right),\\
    C_{202}&=n^0C_{0202}=\frac{1}{\sqrt{1-\frac{2m}{r}}}\left(\frac{m(r-2m)}{r^2}\right),\\
    C_{303}&=n^0C_{0303}=\frac{1}{\sqrt{1-\frac{2m}{r}}}\left(\frac{(r-2m)m \sin^2{\theta}}{r^2}\right),\\
\end{aligned}
\end{equation}
so there are only non-zero temporal 3-form terms. We can then write
\begin{equation}
\begin{gathered}
    C_{abc}=0.
\end{gathered}
\end{equation}
However, we can contract and find the lowered 2-form by contracting the temporal 3-form terms again: $n^0C_{a0b}$
\begin{equation}
\begin{aligned}
    C_{11}&=n^0C_{101}=\frac{1}{{1-\frac{2m}{r}}}\left(-\frac{2m}{r^3}\right)=-\frac{2m}{r^2\left(r-{2m}\right)},\\
    C_{22}&=n^0C_{202}=\frac{1}{{1-\frac{2m}{r}}}\left(\frac{m(r-2m)}{r^2}\right)=\frac{m}{r},\\
    C_{33}&=n^0C_{303}=\frac{1}{{1-\frac{2m}{r}}}\left(\frac{(r-2m)m \sin^2{\theta}}{r^2}\right)=\frac{m \sin^2{\theta}}{r}. \label{lowered2form}
\end{aligned}
\end{equation}
\subsection{Raised 2-form}
The contravariant 2-form of the conformal tensor is written
\begin{equation}
    C^{\alpha\beta}=e^{\alpha\mu}e^{\beta\nu}C_{\mu\nu}.
\end{equation}
There are only the terms 
\begin{equation}
\begin{gathered}
    C^{11}=e^{11}e^{11}C_{11}, \quad
    C^{22}=e^{22}e^{22}C_{22}, \quad
    C^{33}=e^{33}e^{33}C_{33},  \label{raised2form}
\end{gathered}
\end{equation}
and substituting \eqref{spatialterms} and \eqref{lowered2form} into \eqref{raised2form} gives
\begin{equation}
\begin{aligned}
    C^{11}&=\left(1-\frac{2m}{r}\right)\left(1-\frac{2m}{r}\right)\left(-\frac{2m}{r^2\left(r-{2m}\right)}\right)=-\frac{2m\left(r-2m\right)}{r^4}, \\
    C^{22}&=\frac{1}{r^2}\frac{1}{r^2}\frac{m}{r}=\frac{m}{r^5}, \\
    C^{33}&=\frac{1}{r^2 \sin^2{\theta}}\frac{1}{r^2 \sin^2{\theta}}\frac{m \sin^2{\theta}}{r}=\frac{m}{r^5\sin^2{\theta}}. \\
\end{aligned}
\end{equation}

\subsection{First Weyl Scalar $W_1$}

To calculate $C_{ab}C^{ab}$, 
\begin{equation}
\begin{aligned}
    C_{ab}C^{ab}&=C_{11}C^{11}+C_{22}C^{22}+C_{33}C^{33}, \\
    &=\left(-\frac{2m}{r^2\left(r-{2m}\right)}\right)\left(-\frac{2m\left(r-2m\right)}{r^4}\right)+\frac{m}{r}\frac{m}{r^5}+\left(\frac{m \sin^2{\theta}}{r}\right)\frac{m}{r^5\sin^2{\theta}}, \\
    &=\frac{4m^2}{r^6}+\frac{m^2}{r^6}+\frac{m^2}{r^6}, \\
    &=\frac{6m^2}{r^6}. 
\end{aligned}
\end{equation}
Then, $4C_{ab}C^{ab}=\frac{24m^2}{r^6}$, and with $C_{abcd}C^{abcd}=\frac{24m^2}{r^6}$,
\begin{equation}
\begin{gathered}
    W^1=C_{abcd}C^{abcd}+4C_{ab}C^{ab}=\frac{48m^2}{r^6}.
\end{gathered}
\end{equation}

\subsection{Third Weyl Scalar $W^3$}
Following the procedure of the previous Scalar, the third Weyl Scalar is found to be
\begin{equation}
    W^3=W^3_{(0)}+W^3_{(3)}=\frac{48m^3}{r^9}+\frac{48m^3}{r^9}=\frac{96m^3}{r^9},
\end{equation}
or simply put, the only non-independent scalar term is 
\begin{equation}
    W^3=\frac{\left(W^1\right)^{\frac{3}{2}
    }}{2\sqrt{3}}.
\end{equation}

\subsection{Dependencies of Weyl Scalars $W_1$ and $W_2$}
The second and fourth Weyl Scalars require the lapse and scalar densities:
\begin{equation}
\begin{gathered}
    Nt=\sqrt{-^{(4)}g}, \quad 
    t=\sqrt{^{(3)}g}, \quad 
    T^{-1}=t
\end{gathered}
\end{equation}
The value $t$ is simply the determinant of the lowered spatial metric
\begin{equation}
\begin{gathered}
    t=\sqrt{\frac{r^5\sin^2{\theta}}{r-2m}}, \quad 
    T=\sqrt{\frac{r-2m}{r^5\sin^2{\theta}}}. \\
\end{gathered}
\end{equation}
The second Weyl Scalar is dependent on $C_{abc}$ which is zero so this scalar vanishes. Similarly for $W^4$, all of the terms are dependent on $C_{abc}$ with the exception of one being dependent on $C^{cde}$ which is also zero if raised by the contravariant metric.

A heuristic interpretation of the 2-form is as the area-flux through the metric manifold. Similarly, the 4-form may be interpreted as the volume-curvature contraction. Following this, our decomposition suggests that one-half of the intrinsic curvature may be interpreted as self-energy of the geometry, and the other half corresponding to the curvature flux normal to the manifold. This is similar to how the Carminati-McLenaghan invariants decompose into quadratic and cubic contractions.

The non-trivial Schwarzschild BK-Weyl Scalars are: $W^3=\frac{\left(W^1\right)^{\frac{3}{2}}}{2\sqrt{3}}=\frac{96m^3}{r^9}$ which is similar to the Carminati-McLenaghan invariants for the Schwarzschild solution which are $w_2=\frac{(w_1)^{3/2}}{\sqrt{6}}=\frac{6m^3}{r^9}$.

For the Schwarzschild example, there is no interesting information to gleam from the extrinsic curvature, $K_{ab}$ and the conjugate momenta $p^{cd}$ as both of these matrices are zero. This choice of a static solution is slice-dependent which causes the extrinsic curvature and conjugate momenta to vanish. In contrast, for spacetimes with intrinsic rotation, such as the Kerr solution, where these quantities are nontrivial, there may be significant physical insight from comparing the intrinsic and extrinsic curvature for dynamical behavior.

\subsection{Intrinsic vs Extrinsic Foliation Slicing}

Although our Scwarzschild example is static, we can still investigate how the curvature differs on how we foliate our slicing. Since the curvature is solely intrinsic for the Schwarzschild solution the extrinsic curvature is zero. Instead of static slicing in which there is no shift, let us look at the Gullstrand-Painlev\'e \cite{gullstrand1922allgemeine,painleve1921mecanique} slicing in which the lapse is unitary but there is non-zero radial shift
\begin{equation}
\begin{aligned}
    ds^2=-dt^2+\left(dr+\sqrt{2m/r}dt\right)^2+r^2\left(d\theta^2+\sin^2\theta d\phi^2\right).
\end{aligned}
\end{equation}
With $N=1$ we find that $g_{tr}\neq0$, so that $N_r=g_{tr}=\sqrt{\frac{2m}{r}}$. This gives
\begin{equation}
\begin{aligned}
    \Gamma^r_{\theta\theta}=-r, \quad \Gamma^r_{\phi\phi}=-r\sin^2\theta, \quad \Gamma^\theta_{r\theta}=\Gamma^\phi_{r\phi}=\frac{1}{r}
\end{aligned}
\end{equation}
to be used in the covariant derivative
\begin{equation}
    D_a N_b=\partial_a N_b-\Gamma^c_{ab}N_c.
\end{equation}
Similarly to the Schwarzschild slicing, the solution is static in which $\partial_t\gamma_{ab}=0$. The extrinsic curvature, in terms of the covariant derivative, is 
\begin{equation}
\begin{aligned}
    K_{ab}=-\frac{1}{2N}\left(\partial_t \gamma_{ab}-D_a N_b-D_b N_a\right),
\end{aligned}
\end{equation}
where for Gullstrand-Painlev\'e it simplifies to 
\begin{equation}
\begin{aligned}
    K_{ab}=\frac{1}{2}\left(D_a N_b+D_b N_a\right).
\end{aligned}
\end{equation}
This gives
\begin{equation}
    K_{rr}=-\frac{1}{2}\sqrt{\frac{2m}{r^3}}, \quad K_{\theta\theta}=\sqrt{2mr}, \quad K_{\phi\phi}=\sqrt{2mr}\sin^2\theta,
\end{equation}
and from $K=\gamma^{ab}K_{ab}$ we find 
\begin{equation}
    K=K^r_r+K^\theta_\theta+K^\phi_\phi=\frac{3}{2}\sqrt{\frac{2m}{r^3}}.
\end{equation}
Fully raised, $K^{ab}=\gamma^{ac}\gamma^{bd}K_{cd}$, gives
\begin{equation}
\begin{aligned}
    K^{rr}=-\frac{1}{2}\sqrt{\frac{2m}{r^3}}, \quad K^{\theta\theta}=\sqrt{\frac{2m}{r^7}}, \quad K^{\phi\phi}=\sqrt{\frac{2m}{r^7}}\frac{1}{\sin^2\theta}.
\end{aligned}
\end{equation}
From the conjugate momenta equations given in \eqref{conjmomenta} we find
\begin{equation}
\begin{aligned}
    p^{rr}=-2\sin\theta\sqrt{2mr}, \quad p^{\theta\theta}=-\sin\theta\sqrt{\frac{m}{2r^3}}, \quad p^{\phi\phi}=-\frac{1}{\sin\theta}\sqrt{\frac{m}{2r^3}}, \\
    p_{rr}=-2\sin\theta\sqrt{2mr}, \quad p_{\theta\theta}=-\sin\theta\sqrt{\frac{mr}{2}}, \quad p^{\phi\phi}=-{\sin^3\theta}\sqrt{\frac{mr}{2}}. \\
\end{aligned}
\end{equation}
The trace-density relation then agrees giving 
\begin{equation}
    p:=\gamma_{ab}p^{ab}=-2\sqrt{\gamma}K=-3\sqrt{2mr}\sin\theta.
\end{equation}
Now we wish to write the lowered spatial conformal tensor in terms of the canonical variables. We start by writing the relation of the purely spatial Weyl tensor
\begin{equation}
    C_{abcd}:=\gamma_a^\alpha \gamma_b^\beta \gamma_c^\gamma \gamma_d^\delta C_{\alpha\beta\gamma\delta}, \label{spatialweyl}
\end{equation}
and the Riemann tensor 
\begin{equation}
    \gamma_a^\alpha \gamma_b^\beta \gamma_c^\gamma \gamma_d^\delta {}^{(4)}R_{\alpha\beta\gamma\delta}={}^{(3)}R_{abcd}+K_{ac}K_{bd}-K_{ad}K_{bc}. \label{spatialriemann}
\end{equation}
We then want the projected tangent to be preserved so that
\begin{equation}
    \gamma_a^\alpha \gamma_b^\beta \gamma_c^\gamma \gamma_d^\delta g_{\alpha\delta}{}^{(4)}R_{\beta\gamma}=\gamma_{ad}\left(\gamma_b^\beta \gamma_c^\gamma {}^{(4)}R_{\beta\gamma}\right)=\gamma_{ad} {}^{(4)}R_{bd}^\perp,
\end{equation}
where ``$\perp$" indicates the projected tangent terms. For all of the Ricci terms we obtain the antisymmetrized combination
\begin{equation}
    \gamma_{a[d} {}^{(4)}R_{c]b}^\perp-\gamma_{b[d} {}^{(4)}R_{c]a}^\perp.
\end{equation}
For the scalar-curvature terms, we write 
\begin{equation}
    \gamma_a^\alpha \gamma_b^\beta \gamma_c^\gamma \gamma_d^\delta \left(g_{\alpha\gamma}g_{\beta\delta}-g_{\alpha\delta}g_{\beta\gamma}\right){}^{(4)} R=\left(\gamma_{ac}\gamma_{bd}-\gamma_{ad}\gamma_{bc}\right){}^{(4)}R.
\end{equation}
We define the contracted Gauss relation in terms of the tangential projection of the 4D Ricci tensor as
\begin{equation}
    S_{ab}:={}^{(4)}R_{ab}^\perp={}^{(3)}R_{ab}+K K_{ab}-K_a^iK_{bi},  \label{sij}
\end{equation}
and its trace
\begin{equation}
    S:={}^{(4)}R^\perp:=\gamma^{ab}{}^{(4)}R_{ab}^\perp={}^{(3)}R+K^2-K_{ij}K^{ij} .\label{s}
\end{equation}

The conformal tensor \eqref{conformal tensor}, may be written using \eqref{spatialweyl}, \eqref{spatialriemann}, \eqref{sij}, and \eqref{s} as
\begin{equation}
\begin{aligned}
    C_{abcd}=&{}^{(3)}R_{abcd}+K_{ac}K_{bd}-K_{ad}K_{bc} \\
    &-\frac{1}{2}\left(\gamma_{a[c}S_{d]b}-\gamma_{b[c}S_{d]a}\right)+\frac{1}{6}S\gamma_{a[c}\gamma_{d]b}.
\end{aligned} \label{conformalagain}
\end{equation}
Writing everything purely in canonical variables we get
\begin{equation}
\begin{aligned}
    K^{ab}=\frac{1}{\sqrt{\gamma}}\left(p^{ab}-\frac{\gamma^{ab}p}{2}\right), \quad K_{ab}=\frac{1}{\sqrt{\gamma}}\left(p_{ab}-\frac{\gamma_{ab}p}{2}\right), \quad K=-\frac{p}{2\sqrt{\gamma}}, \label{canonicalextrinsic}
\end{aligned}
\end{equation}
so that \eqref{sij} and \eqref{s} become
\begin{equation}
\begin{aligned}
    S_{ab}=&{}^{(3)}R_{ab}-\frac{1}{\gamma}p_a^i p_{ib}+\frac{1}{2\gamma}p p_{ab}, 
\end{aligned} \label{ses}
and
\end{equation}
\begin{equation}
    S={}^{(3)}R-\frac{1}{\gamma}p_{ij} p^{ij}+\frac{1}{2\gamma}p^2, \label{ses2}
\end{equation}
respectively. Using \eqref{canonicalextrinsic}, \eqref{ses}, and \eqref{ses2} we write \eqref{conformalagain} as 
\begin{equation}
\begin{aligned}
    C_{abcd}=&{}^{(3)}R_{abcd}+\frac{1}{\gamma}\left[\left(p_{ac}-\frac{1}{2}\gamma_{ac}p\right)\left(p_{bd}-\frac{1}{2}\gamma_{bd}p\right)-\left(p_{ad}-\frac{1}{2}\gamma_{ad}p\right)\left(p_{bc}-\frac{1}{2}\gamma_{bc}p\right)\right] \\
    &-\frac{1}{2}\left(\gamma_{a[c}S_{d]b}-\gamma_{b[c}S_{d]a}\right)+\frac{1}{6}S\gamma_{a[c}\gamma_{d]b}.
\end{aligned} \label{conformalcanonical}
\end{equation}
Equation \eqref{conformalcanonical} is the spatial Conformal tensor in terms of the canonical phase space variables. 

When comparing the two metrics, Schwarzschild and Gullstrand-Painlev\'e, they only differ by the foliation of the static slicing by being described by the intrinsic or extrinsic curvature without altering the 4D information. The tidal field, or the gravitational behavior once the gauge is chosen, is an intrinsic property of the spacetime and is independent of the coordinate representation or the instantaneous labeling of points within the chosen frame.

We emphasize that it is the bookkeeping that differs: in the Schwarzschild static slicing the curvature sits in ${}^{(3)}R_{abcd}$ with $p_{ab}=0$, while for Gullstrand-Painlev\'e slicing ${}^{(3)}R_{abcd}=0$ and the same tidal information sits in $p_{ab}\neq0$. The 4D curvature is unchanged. This illustrates that gravity is not a fundamental force but more better interpreted as the geodesic deviation relative to two observers. Apparent forces such as Coriolis or centrifugal effects reflect the kinematics of the chosen coordinates rather than genuine physical interactions. This indicates that when working with the Hamiltonian, the tidal gravity can be described by either the configuration space or from the momentum space depending on the foliation choice. Intermediate slicings, such as those interpolating between Schwarzschild and Gullstrand-Painlev\'e coordinates, can be formalized by introducing a one-parameter family of canonical transformations, continuously mixing configuration and momentum representation of the same curvature content. 

Thus an ICRF distinguishes gauges by the same invariant tidal eigen-structures as is done in characterization. If an affect dissappears when working with a proper ICRF, it does not alter the Weyl eigenvalues or the null principal direction, it is due to the gauge. For example in the Schwarzschild solution the coordinate chart singularity is at the event horizon ($r=2M$), where the Horizon remains regular for Gullstrand-Painlev\'e. If the effect remains under all smooth coordinate transformations, it represents an intrinsic tidal or geometric field and not a coordinate artifact. Observer-dependent effects, must be treated separately (and carefully). 

The ICRF identifies a preferred geometrical basis from the curvature invariance, be it Petrov classification or the eigenframe of the Weyl tensor. The use of an ICRF provides an operational means of implementing diffeomorphism invariance, allowing one to distinguish genuine curvature from inertial artifacts, though it does not single out a unique physical frame. In this sense, the ICRF is analogous to other invariant constructions, \emph{e.g.} Segre or Petrov canonical frames, or frames defined by curvature Killing vectors, which provide distinct but equally valid realizations of diffeomorphism-invariant reference structures.

\section{Intrinsic Coordinate Reference Frame}

Since the Weyl scalars presented here are solved in terms of the canonical phase-space variables, we may choose four independent phase-space scalar functions to describe an ICRF. These scalar functions may be chosen as our imposed gauge conditions, so that we may choose an intrinsic time parameter (or evolutionary parameter) that depends on the phase-space variables \cite{komar_construction_1958, bergmann_1961}. For the generic asymmetric case of an ICRF, we assume the scalars are independent so we may follow the procedure given in \cite{pons_2005}. 

We assume a local metric $g^{\mu\nu}(x)$ of a given gauge orbit is given in terms of a set of coordinates $x^{\mu}$, we introduce another local metric of the same gauge orbit by $g'^{\mu\nu}(x')$ defined by $x'^{\mu}$. We also assume our metrics are related by an active diffeomorphism generated by the infinitesimal vector field $\epsilon^{\mu}{\partial}_{\mu}$, written
\begin{equation}
    g'^{\mu\nu}(x')=g^{\mu\nu}(x)+\pounds_{\epsilon^{\mu}{\partial}_{\mu}}\left(g^{\mu\nu}(x)\right),  \\ \label{diffeomorphism_relation}
\end{equation}
where $\pounds_{\epsilon^{\mu}{\partial}_{\mu}}$ represents the Lie derivative with respect to the vector $\epsilon^{\mu}{\partial}_{\mu}$. Using the independence of our Weyl scalars allows us to define four scalar functions, $A^I(x)$ (with $A'^I(x')$ corresponding to $g'^{\mu\nu}(x')$), to imply
\begin{equation}
    \det\left(\frac{{\partial}A^I(x)}{{\partial}x^{\mu}}\right)\neq0. \\
\end{equation}
The relation \eqref{diffeomorphism_relation}, states $A^I(x)=A'^I(x')$, indicating that the scalar functions dictate a metric-dependent change of coordinates. We adopt these four scalar functions to describe an ICRF
\begin{equation}
    X^I:=A^I, \label{DRF}
\end{equation}
which is often referred to as ``intrinsic coordinates". This ICRF is dependent on the background fields of the intrinsic geometry, and a set of satellites measuring these scalars could be used to describe the local four-geometry analogous to that of Global Positioning Systems (GPS) as proposed by Rovelli \cite{rovelli_2002a}. 

We show also that if we describe this ICRF as four spacetime ``coordinates'' $x^{\mu}$, by invoking the gauge conditions
\begin{equation}
    x^{\mu}=X^{\mu}[g_{ab},p^{cd}], \label{gauge condition}
\end{equation}
we can allow characterization of distinct solutions to Einstein's field equations (EFEs). With the gauge condition \eqref{gauge condition}, any phase-space solution for EFEs, along with the complete generator \eqref{generator} will allow transforming the given solution back into the form of \eqref{gauge condition}. 

This implies that the use of \eqref{gauge condition} would lead to a complete and non-redundant set of observables that relate the generators and the infinitesimal canonical transformations, and identify equivalence classes of solutions of EFEs \cite{bergmann_1962,pons_1997}. A change of any of the initial coordinates (gauge-constraints) would result in changing the gauge-orbit and consequently the physical state \cite{bergmann_1962,pons_1997,rovelli_2002b,pons_2005,salisbury_cartan_2022}. Transforming to this ICRF will yield identical metric functions \cite{pons_2005}. Using a fixed ICRF we get physically distinct solutions with a change in the independent phase-space variables for a given initial intrinsic time. In other words, each distinct solution represents a full equivalence class under general (non-phase-space dependent) coordinate transformations.

\section{Conclusion}

In this paper we present the novel phase-space solution of the set of Weyl scalars in tensorial form, two of which were derived by Bergmann and Komar in 1960 \cite{bk_1960}. Although the nature of a full diffeomorphism covariance was not fully understood at the time, Bergmann and Komar undertook a Hamilton-Jacobi approach to GR and ultimately showed how classical solutions of the Hamilton-Jacobi equations could be interpreted as identifying the reduced phase-space that results from the quotienting of the full 4D diffeomorphism group. In \cite{komar_construction_1958}, Komar suggested introducing an ICRF through use of the Weyl Scalars, and so it is odd that both Bergmann and Komar never employed the Weyl scalars while they worked closely with their Hamilton-Jacobi analysis \cite{bk_1960}.

The implications of these Weyl scalars cannot be emphasized enough since the implementation of such a dynamical coordinate system can be used to solve for local observables \cite{chataignier_diss_2022}. Another route is using the ICRF in a perturbational approach to an isotropic and homogeneous toy model of the universe, e.g.~the Friedmann-Lema\^itre-Robertson-Walker  metric, to approximate (weakly) a set of Dirac observables \cite{salisbury_2023}. It is also of great interest to compare the Weyl scalars to that of the SPCIs such as the Cartan-Karlhede or Carminati-McLenaghan invariants which have been invaluable for characterizing spacetimes such as through Petrov classification \cite{karlhede1980review,carminati1991algebraic,zakhary_2003}. These Weyl scalars (or an ICRF) may also aid with the development of recent extended quantum theories  \cite{bender_1999,bender_2002,bender_2003,mostafazadeh_2004,mostafazadeh_2007,mostafazadeh_2018,mostafazadeh_2020,kabel_2024}, especially in those that stress the non-triviality of quantum reference frames \cite{kabel_2024}. 

The non-physicality of the Weyl scalars is far from aesthetically pleasing but is expected to be invaluable with the works of numerical relativity. It is of interest to compute the Weyl scalars for known solutions of EFEs such as the Schwarzschild, Kerr, Reissner-Nordstr\"om, and Kerr-Newman metrics to investigate the background fields for an ICRF in `Geometrodynamics'. It is also of interest for the applications of warp-drive and binary merger simulations. We plan to investigate the spinorial construction of the Weyl scalars to be compared to traditional coordinate-invariants in a follow-up paper, and how to distinguish individual internal parameters such as an evolutionary parameter that may be interpreted as ``time''.

\newpage

\bibliographystyle{unsrt-phys-eucos}
\bibliography{ref}

\appendix

\section{Generating Terms}

Here we give an example as to how the product of \eqref{32a$'$} is found, and then list the other products for \eqref{W3}. Starting from \eqref{G} where we substitute \eqref{projection tensor} so that
\begin{equation}
\begin{gathered}
    G^{\mathcal{BC}}=(e^{\gamma\rho}-n^{\gamma}n^{\rho})(e^{\delta\sigma}-n^{\delta}n^{\sigma})\\
    -(e^{\delta\rho}-n^{\delta}n^{\rho})(e^{\gamma\sigma}-n^{\gamma}n^{\sigma}), \notag
\end{gathered}
\end{equation}
and expanding gives
\begin{equation}
\begin{gathered}
    G^{\mathcal{BC}}=e^{\gamma\rho}e^{\delta\sigma}-e^{\delta\rho}e^{\gamma\sigma}-e^{\gamma\rho}n^{\delta}n^{\sigma}-e^{\delta\sigma}n^{\gamma}n^{\rho}+e^{\delta\rho}n^{\gamma}n^{\sigma}+e^{\gamma\sigma}n^{\delta}n^{\rho}+n^{\gamma}n^{\rho}n^{\delta}n^{\sigma}-n^{\delta}n^{\rho}n^{\gamma}n^{\sigma}.  \notag
\end{gathered}
\end{equation}
Similarly
\begin{equation}
    \begin{gathered}
        G^{\mathcal{DA}}=e^{\mu\alpha}e^{\nu\beta}-e^{\nu\alpha}e^{\mu\beta}-e^{\mu\alpha}n^{\nu}n^{\beta}-e^{\nu\beta}n^{\mu}n^{\alpha}+e^{\nu\alpha}n^{\mu}n^{\beta}+e^{\mu\beta}n^{\nu}n^{\alpha}+n^{\mu}n^{\alpha}n^{\nu}n^{\beta}-n^{\nu}n^{\alpha}n^{\mu}n^{\beta}.  \notag
    \end{gathered}
\end{equation}
So the product of the two lines above is
\begin{equation}
\begin{gathered}
    G^{\mathcal{BC}}G^{\mathcal{DA}}=e^{\gamma\rho}e^{\delta\sigma}e^{\mu\alpha}e^{\nu\beta}-e^{\gamma\rho}e^{\delta\sigma}e^{\nu\alpha}e^{\mu\beta}-e^{\delta\rho}e^{\gamma\sigma}e^{\mu\alpha}e^{\nu\beta}+e^{\delta\rho}e^{\gamma\sigma}e^{\nu\alpha}e^{\mu\beta} \\
    -e^{\gamma\rho}e^{\delta\sigma}e^{\mu\alpha}n^{\nu}n^{\beta}-e^{\gamma\rho}e^{\delta\sigma}e^{\nu\beta}n^{\mu}n^{\alpha}+e^{\gamma\rho}e^{\delta\sigma}e^{\nu\alpha}n^{\mu}n^{\beta}+e^{\gamma\rho}e^{\delta\sigma}e^{\mu\beta}n^{\nu}n^{\alpha}\\
    +e^{\delta\rho}e^{\gamma\sigma}e^{\mu\alpha}n^{\nu}n^{\beta}+e^{\delta\rho}e^{\gamma\sigma}e^{\nu\beta}n^{\mu}n^{\alpha}-e^{\delta\rho}e^{\gamma\sigma}e^{\nu\alpha}n^{\mu}n^{\beta}-e^{\delta\rho}e^{\gamma\sigma}e^{\mu\beta}n^{\nu}n^{\alpha}\\
    -e^{\mu\alpha}e^{\nu\beta}e^{\gamma\rho}n^{\delta}n^{\sigma}-e^{\mu\alpha}e^{\nu\beta}e^{\delta\sigma}n^{\gamma}n^{\rho}+e^{\mu\alpha}e^{\nu\beta}e^{\delta\rho}n^{\gamma}n^{\sigma}+e^{\mu\alpha}e^{\nu\beta}e^{\gamma\sigma}n^{\delta}n^{\rho}\\
    +e^{\nu\alpha}e^{\mu\beta}e^{\gamma\rho}n^{\delta}n^{\sigma}+e^{\nu\alpha}e^{\mu\beta}e^{\delta\sigma}n^{\gamma}n^{\rho}-e^{\nu\alpha}e^{\mu\beta}e^{\delta\rho}n^{\gamma}n^{\sigma}-e^{\nu\alpha}e^{\mu\beta}e^{\gamma\sigma}n^{\delta}n^{\rho} \\
    +e^{\gamma\rho}e^{\mu\alpha}n^{\delta}n^{\sigma}n^{\nu}n^{\beta}+e^{\gamma\rho}e^{\nu\beta}n^{\delta}n^{\sigma}n^{\mu}n^{\alpha}-e^{\gamma\rho}e^{\nu\alpha}n^{\delta}n^{\sigma}n^{\mu}n^{\beta}-e^{\gamma\rho}e^{\mu\beta}n^{\delta}n^{\sigma}n^{\nu}n^{\alpha}\\
    +e^{\delta\sigma}e^{\mu\alpha}n^{\gamma}n^{\rho}n^{\nu}n^{\beta}+e^{\delta\sigma}e^{\nu\beta}n^{\gamma}n^{\rho}n^{\mu}n^{\alpha}-e^{\delta\sigma}e^{\nu\alpha}n^{\gamma}n^{\rho}n^{\mu}n^{\beta}-e^{\delta\sigma}e^{\mu\beta}n^{\gamma}n^{\rho}n^{\nu}n^{\alpha}\\
    -e^{\delta\rho}e^{\mu\alpha}n^{\gamma}n^{\sigma}n^{\nu}n^{\beta}-e^{\delta\rho}e^{\nu\beta}n^{\gamma}n^{\sigma}n^{\mu}n^{\alpha}+e^{\delta\rho}e^{\nu\alpha}n^{\gamma}n^{\sigma}n^{\mu}n^{\beta}+e^{\delta\rho}e^{\mu\beta}n^{\gamma}n^{\sigma}n^{\nu}n^{\alpha}\\
    -e^{\gamma\sigma}e^{\mu\alpha}n^{\delta}n^{\rho}n^{\nu}n^{\beta}-e^{\gamma\sigma}e^{\nu\beta}n^{\delta}n^{\rho}n^{\mu}n^{\alpha}+e^{\gamma\sigma}e^{\nu\alpha}n^{\delta}n^{\rho}n^{\mu}n^{\beta}+e^{\gamma\sigma}e^{\mu\beta}n^{\delta}n^{\rho}n^{\nu}n^{\alpha}. \notag
\end{gathered}
\end{equation}
This becomes unwieldy so we decompose it into the form
\begin{equation}
    G^{\mathcal{BC}}G^{\mathcal{DA}}\equiv\left(G^{\mathcal{BC}}G^{\mathcal{DA}}\right)_{(0)}+\left(G^{\mathcal{BC}}G^{\mathcal{DA}}\right)_{(1)}+\left(G^{\mathcal{BC}}G^{\mathcal{DA}}\right)_{(2)}, 
\end{equation}
where we write
    \begin{align*}
    \left(G^{\mathcal{BC}}G^{\mathcal{DA}}\right)_{(0)}{\equiv}&~G^{\mathcal{BC}}G^{\mathcal{DA}}+\mathcal{O}\left(n^{\alpha}n^{\beta}\right), \\ \notag
    \left(G^{\mathcal{BC}}G^{\mathcal{DA}}\right)_{(1)}{\equiv}&~G^{\mathcal{BC}}G^{\mathcal{DA}}-\left(G^{\mathcal{BC}}G^{\mathcal{DA}}\right)_{(0)}
    +\mathcal{O}\left(n^{\alpha}n^{\beta}n^{\gamma}n^{\delta}\right), \\
    \left(G^{\mathcal{BC}}G^{\mathcal{DA}}\right)_{(2)}{\equiv}&~G^{\mathcal{BC}}G^{\mathcal{DA}}-\left(G^{\mathcal{BC}}G^{\mathcal{DA}}\right)_{(0)}
    -\left(G^{\mathcal{BC}}G^{\mathcal{DA}}\right)_{(1)},    \notag
    \end{align*}
until no more terms remain. We use the symmetry
\begin{equation}
    e^{\mu\alpha}e^{\nu\beta}=-e^{\nu\alpha}e^{\mu\beta}, \label{symmetry} 
\end{equation}
to combine like-terms, e.g.
\begin{equation}
\begin{gathered}
     \left(G^{\mathcal{BC}}G^{\mathcal{DA}}\right)_{(0)}\equiv~e^{\gamma\rho}e^{\delta\sigma}e^{\mu\alpha}e^{\nu\beta}-e^{\gamma\rho}e^{\delta\sigma}e^{\nu\alpha}e^{\mu\beta}-e^{\delta\rho}e^{\gamma\sigma}e^{\mu\alpha}e^{\nu\beta}+e^{\delta\rho}e^{\gamma\sigma}e^{\nu\alpha}e^{\mu\beta},  \notag
\end{gathered}
\end{equation}
becomes
\begin{equation}
    \begin{gathered}
        \left(G^{\mathcal{BC}}G^{\mathcal{DA}}\right)_{(0)}=4e^{\gamma\rho}e^{\delta\sigma}e^{\mu\alpha}e^{\nu\beta}. \notag
    \end{gathered}
\end{equation}
This follows for the other values
\begin{align*}
        \left(G^{\mathcal{BC}}G^{\mathcal{DA}}\right)_{(0)}=&~4e^{\gamma\rho}e^{\delta\sigma}e^{\mu\alpha}e^{\nu\beta}, \\ \notag
        \left(G^{\mathcal{BC}}G^{\mathcal{DA}}\right)_{(1)}=&-16e^{\gamma\rho}e^{\delta\sigma}e^{\mu\alpha}n^{\nu}n^{\beta}, \\ \notag
        \left(G^{\mathcal{BC}}G^{\mathcal{DA}}\right)_{(2)}=&~16e^{\gamma\rho}e^{\mu\alpha}n^{\delta}n^{\sigma}n^{\nu}n^{\beta}, \\ \notag
        G^{\mathcal{BC}}_{(0)}=&~2e^{\gamma\rho}e^{\delta\sigma}, \\ \notag
        G^{\mathcal{BC}}_{(1)}=&-2^2e^{\gamma\rho}n^{\delta}n^{\sigma}, \\ \notag
        \left(G^{\mathcal{BC}}G^{\mathcal{DE}}G^{\mathcal{FA}}\right)_{(0)}=& 
        ~2^3e^{\gamma\rho}e^{\delta\sigma}e^{\mu\kappa}e^{\nu\lambda}e^{\tau\alpha}e^{\omega\beta}, \\ \notag
         \left(G^{\mathcal{BC}}G^{\mathcal{DE}}G^{\mathcal{FA}}\right)_{(1)}=&-2^4(e^{\gamma\rho}e^{\delta\sigma}e^{\mu\kappa}e^{\nu\lambda}e^{\tau\alpha}n^{\omega}n^{\beta}+2e^{\gamma\rho}e^{\delta\sigma}e^{\mu\kappa}e^{\tau\alpha}e^{\omega\beta}n^{\nu}n^{\lambda}), \\ \notag
        \left(G^{\mathcal{BC}}G^{\mathcal{DE}}G^{\mathcal{FA}}\right)_{(2)}=&~2^5e^{\gamma\rho}e^{\mu\kappa}e^{\tau\alpha}e^{\omega\beta}n^{\delta}n^{\sigma}n^{\nu}n^{\lambda}+2^6e^{\gamma\rho}e^{\delta\sigma}e^{\mu\kappa}e^{\tau\alpha}n^{\nu}n^{\lambda}n^{\omega}n^{\beta}, \\ \notag
        \left(G^{\mathcal{BC}}G^{\mathcal{DE}}\right)_{(0)}=&-2^2e^{\gamma\rho}e^{\delta\sigma}e^{\mu\kappa}e^{\nu\lambda}, \\ \notag
        \left(G^{\mathcal{BC}}G^{\mathcal{DE}}\right)_{(1)}=&-2^4\left(e^{\gamma\rho}e^{\delta\sigma}e^{\mu\kappa}n^{\nu}n^{\lambda}+e^{\gamma\rho}e^{\mu\kappa}e^{\nu\lambda}n^{\delta}n^{\sigma}\right), \\ \notag
        \left(G^{\mathcal{BC}}G^{\mathcal{DE}}\right)_{(2)}=&-2^5e^{\gamma\rho}e^{\mu\kappa}n^{\delta}n^{\sigma}n^{\nu}n^{\lambda}. \\ \notag
\end{align*}
We recombine our higher order terms so that
\begin{subequations}
    \begin{equation}
        G^{\mathcal{BC}}G^{\mathcal{DA}}=4e^{\gamma\rho}e^{\delta\sigma}e^{\mu\alpha}e^{\nu\beta}-16e^{\gamma\rho}e^{\delta\sigma}e^{\mu\alpha}n^{\nu}n^{\beta}+16e^{\gamma\rho}e^{\mu\alpha}n^{\delta}n^{\sigma}n^{\nu}n^{\beta}, \label{WG1}
    \end{equation}
    \begin{equation}
        G^{\mathcal{BC}}=2e^{\gamma\rho}e^{\delta\sigma}-2^2e^{\gamma\rho}n^{\delta}n^{\sigma}, \label{WG2}
    \end{equation}
    \begin{equation}
        \begin{gathered}
            G^{\mathcal{BC}}G^{\mathcal{DE}}G^{\mathcal{FA}}=2^3e^{\gamma\rho}e^{\delta\sigma}e^{\mu\kappa}e^{\nu\lambda}e^{\tau\alpha}e^{\omega\beta}-2^4(e^{\gamma\rho}e^{\delta\sigma}e^{\mu\kappa}e^{\nu\lambda}e^{\tau\alpha}n^{\omega}n^{\beta}+2e^{\gamma\rho}e^{\delta\sigma}e^{\mu\kappa}e^{\tau\alpha}e^{\omega\beta}n^{\nu}n^{\lambda})\\
            +2^5e^{\gamma\rho}e^{\mu\kappa}e^{\tau\alpha}e^{\omega\beta}n^{\delta}n^{\sigma}n^{\nu}n^{\lambda}+2^6e^{\gamma\rho}e^{\delta\sigma}e^{\mu\kappa}e^{\tau\alpha}n^{\nu}n^{\lambda}n^{\omega}n^{\beta}, \label{WG3}
        \end{gathered}
    \end{equation}
    \begin{equation}
        G^{\mathcal{BC}}G^{\mathcal{DE}}=-2^2e^{\gamma\rho}e^{\delta\sigma}e^{\mu\kappa}e^{\nu\lambda}-2^4\left(e^{\gamma\rho}e^{\delta\sigma}e^{\mu\kappa}n^{\nu}n^{\lambda}+e^{\gamma\rho}e^{\mu\kappa}e^{\nu\lambda}n^{\delta}n^{\sigma}\right)-2^5e^{\gamma\rho}e^{\mu\kappa}n^{\delta}n^{\sigma}n^{\nu}n^{\lambda}. \label{WG4}
    \end{equation}
\end{subequations}

\end{document}